\documentclass[11pt]{JHEP3}

\usepackage{epsfig,multicol,bbm}
\usepackage{cite}
\usepackage{amsmath,amsfonts,amssymb,longtable}
\usepackage{array} 
\usepackage{graphicx}
\usepackage{subfig}
\usepackage{enumerate}
\input epsf.sty

\def\be{\begin{equation}}
\def\ee{\end{equation}}
\def\ba{\begin{array}}
\def\bacc{\begin{array} {cc}}
\def\ea{\end{array}}
\def\bea{\begin{eqnarray}}
\def\eea{\end{eqnarray}}
\def\bd{\begin{displaymath}}
\def\ed{\end{displaymath}}
\def\calN{{\mathcal N}}
\def\calF{{\mathcal F}}

\def\Z{\mathbb Z}

\def\N{{\rm N}}
\def\L{{\rm L}}
\def\M{{\rm M}}

\def\K{{\rm K}}
\def\k{{\bf k}}
\def\mod{{\,\,\rm mod}\,\,}
\def\nn{\nonumber}

\usepackage{cancel, color}

\title{\huge  R-charge Conservation and More in Factorizable and Non-Factorizable Orbifolds}

\author{Nana G.~Cabo Bizet$^{a,b,c}$, Tatsuo Kobayashi$^d$, Dami\'an K.~Mayorga Pe\~na$^a$,   \hskip2cm Susha L.~Parameswaran$^e$, Matthias Schmitz$^{a}$, Ivonne Zavala$^f$ \\ \\
${}^a$Bethe Center for Theoretical Physics and
Physikalisches Institut der Universit\"at Bonn,
Nussallee 12, 53115 Bonn, Germany\\
${}^b$Centro de Aplicaciones Tecnol\'ogicas y Desarrollo Nuclear, Calle 30, esq.a 5ta Ave, Miramar, 6122 La Habana, Cuba\\
${}^c$Theory Group, Physics Department, CERN CH-1211, Gen\`eve 23, Switzerland\\
${}^d$Department of Physics, Kyoto University, Kyoto 606-8502, Japan \\
${}^e$Department of Mathematics and Physics, Leibniz Universit\"at Hannover, Welfengarten 1, 30167 Hannover, Germany\\ 
${}^f$Centre for Theoretical Physics, University of Groningen,
Nijenborgh 4, 9747 AG Groningen, The Netherlands\\
\\
E-mail: {\email{nana@th.physik.uni-bonn.de}
\email{kobayash@gauge.scphys.kyoto-u.ac.jp}\,,
\email{damian@th.physik.uni-bonn.de}\,,
\email{susha.parameswaran@itp.uni-hannover.de}\,, 
\email{mschmitz@th.physik.uni-bonn.de}\,,
\email{e.i.zavala@rug.nl}}
}

\preprint{CERN-PH-TH/2012-365\\KUNS-2431}

\abstract{We consider the string theory origin of R-charge conservation laws in heterotic orbifold compactifications, deriving the corresponding string coupling selection rule for factorizable and non-factorizable orbifolds, with prime ordered and non-prime ordered point groups.  R-charge conservation arises due to symmetries among the worldsheet instantons that can mediate the couplings.  Among our results is a previously missed non-trivial contribution to the conserved R-charges from the $\gamma$-phases in non-prime orbifolds, which weakens the R-charge selection rule.  Symmetries among the worldsheet instantons can also lead to additional selection rules for some couplings.  We make a similar analysis for Rule 4 or the ``torus lattice selection rule''.  Moreover, we identify a new string selection rule, that we call Rule 6 or the ``coset vector selection rule''.}


\begin{document}

\section{Introduction}
In recent years remarkable progress has been made in the quest for a string theoretic description of the standard model of particle physics.  In particular we now have hundreds of explicit models whose low energy matter spectra are potentially realistic -- containing the standard model particles 
and no chiral exotics -- in several classes of string constructions; heterotic smooth Calabi-Yau compactifications \cite{smoothCY}, heterotic orbifold compactifications \cite{hetorbs}, Gepner models \cite{gepner}, free fermionic constructions \cite{freefermionic}, D-brane models \cite{Dbranes}, F-theory \cite{Ftheory} (see \cite{raby,Faraggi:2010fi,GatoRivera:2010gv,eran} for some recent reviews). It is now essential to go beyond the particle content, and consider the low energy effective field theory that describes their phenomenology.  Toroidal orbifold compactifications are attractive in this respect, as the corresponding action can at least in principle be computed explicitly, via the string CFT \cite{Dixon,HV}, which is free.

A first question one asks is which couplings in the action are allowed to be non-vanishing.  The topic of selection rules for non-vanishing superpotential couplings in heterotic orbifold compactifications was revisited recently in \cite{KPRZ}.  By studying the corresponding L-point correlation functions, an apparently forgotten rule \cite{Rule4} was identified, and a new rule discovered.  These Rules 4 and 5 arise from the properties of the worldsheet instantons that can mediate couplings.

We seek in the present paper to understand the stringy nature of R-symmetries in heterotic orbifold compactifications.  It has generally been assumed that this issue is well understood for factorizable orbifolds \cite{Rcharge&miracle, Rcharge}, but not understood at all for non-factorizable orbifolds \cite{patrick, saulsthesis}.  We provide a derivation of the R-charge conservation law from the orbifold CFT, for factorizable and non-factorizable orbifolds, with prime ordered and non-prime ordered twists.  
We restrict our present analysis to compactifications without discrete Wilson lines.  This is because the effect of discrete Wilson lines on the string couplings has not yet been worked out, since a consistent action description in this case is not known \cite{Erler:1991nr, Kobayashi:2003vi}.   

We show that the R-charge conservation law also emerges due to properties of the worldsheet instantons, with symmetries in the orbifold geometry leading to symmetries amongst the worldsheet instantons wrapping the orbifold space.
For non-factorizable orbifolds, we find charge conservation laws which correspond to non-R symmetries, as well as some R-charge selection rules.
At the same time, we also encounter a new stringy rule, which we may call Rule 6, which applies in some orbifolds to some couplings.  Its origin is similar to that of charge conservation and Rule 4, arising due to symmetries among worldsheet instantons.

Our results on the R-charge conservation agree with the current literature for the prime factorizable orbifolds, but for the non-prime case we find a non-trivial contribution to the conserved R-charges from the $\gamma$-phase, which has previously been unnoted.  
As a consequence, some couplings that are forbidden by the old R-charge conservation are actually non-vanishing. 
This is particularly important, as a favourite choice for Orbifolders constructing potentially realistic orbifold compactifications \cite{hetorbs,mssm,Rcharge} has been non-prime orbifolds like $T^6/\Z_{6-II}$. 
Another observation that differs from the current picture is that factorizable orbifolds do not always lead to three independent R-charge conservation laws, with one associated to each plane.  In fact, planes for which the orbifold twist is non-prime turn out to all contribute to a single R-charge conservation law.

The paper is organized as follows.  In Section 2, we provide a brief review of the orbifold setup and outline how string selection rules can be derived from the string correlation functions.  We build upon this discussion in Section 3, first considering the discrete 
Lorentz symmetries enjoyed by the orbifold geometry, and then deriving directly from the CFT the charge conservation laws and Rule 6 for factorizable and non-factorizable orbifolds.  In Section 4, we turn to Rule 4.  Finally, in Section 5, we provide a summary of our results, and discuss their significance.  In an appendix we give a more detailed classification of orbifold automorphisms, for future reference.

\section{Orbifold CFT Review}

Let us begin by briefly describing the orbifold geometry and associated conformal field theory \cite{Dixon, HV}, which provides the setting for our discussion.  In this section, we introduce the orbifold and the string states that emerge when heterotic string theory is compactified on the orbifold.  Then we describe the basic ingredients of the corresponding worldsheet conformal field theory, in particular the vertex operators and correlation functions.  Finally, we review how the correlation functions can be used to derive string coupling selection rules, referring to \cite{KPRZ} for more details.

\subsection{Orbifold geometry, the space group and conjugacy classes}

A $\Z_\N$ orbifold is constructed by dividing ${\mathbb R}^6$ by a six-dimensional lattice $\Lambda$ in order to obtain a torus, and subsequently modding out by some $\Z_\N$ automorphism.  The $\Z_\N$ automorphism is called the point group $P$, and defining the space group $S$ as $S = \Lambda \rtimes P$, we can thus describe the orbifold as $T^6/P$ or ${\mathbb R}^6/S$ \cite{ZN}.

We call the generator of $P$ the twist, $\theta$, and it is convenient to diagonalize it, using complex coordinates to describe the torus and writing $\theta$ as:
\be
\theta = \textrm{diag}\left(e^{2\pi\imath v_1}, e^{2\pi\imath v_2},e^{2\pi\imath v_3} \right) \,.
\ee
The vector $v=(v_1,v_2,v_3)$ is referred to as the twist vector. The orbifold geometry has a number of special points that are fixed under the action of the point group and torus lattice shifts.  We denote with $f^{(k)}$ a point fixed under $\theta^k$, that is $f^{(k)} = \theta^k f^{(k)} + \lambda$, for some $\lambda \in \Lambda$.  Thus we can associate to each fixed point a space group element $g=(\theta^k,\lambda)$ or $g=(\theta^k,(1-\theta^k)f^{(k)})$.  

We consider compactifications of the heterotic string theory on this orbifold geometry.  The requirement of ${\cal N}=1$ supersymmetry in four dimensions then restricts the $\Z_\N$ point group to nine possibilities\footnote{In addition there are nine $\Z_\N \times \Z_\M$ orbifolds, for which we could make an analogous discussion.  More sophisticated orbifold constructions have also been considered for instance in Ref. \cite{fischer}.} listed in Table \ref{T:ZN}.  
\TABLE[t]{

\centering
\begin{tabular}{|c|c|} \hline
Point group & Twist vector \\  \hline \hline
$\Z_3$ & $(1,1,-2)/3$  \\
$\Z_4$ & $(1,1,-2)/4$  \\
$\Z_{6-I}$ & $(1,1,-2)/6$  \\
$\Z_{6-II}$ & $(1,2,-3)/6$  \\
$\Z_7$ & $(1,2,-3)/7$ \\
$\Z_{8-I}$ & $(2,1,-3)/8$  \\
$\Z_{8-II}$ & $(1,3,-4)/8$  \\
$\Z_{12-I}$ & $(4,1,-5)/12$  \\
$\Z_{12-II}$ & $(1,5,-6)/12$  \\ \hline
\end{tabular}

\caption{All $\Z_N$ point groups and\\
twist vectors for 6D orbifolds lea-\\
ding to ${\cal N}=1$ SUSY \cite{ZN}.}
\label{T:ZN}

}

We will see that the properties of the orbifold compactification depend strongly on whether the order $\N$ of the point group is prime or non-prime.  Given a point group, several underlying torus lattices may be possible \cite{lattices, Kobayashi:1991rp,CGM,fischer}, the only requirement being that the group of lattice automorphisms contains $\Z_\N$.  Orbifolds may then also be classed according to whether or not the underlying torus lattice factorizes into three orthogonal complex planes.  If so, the orbifold is called factorizable, although this is rather a
misnomer as it is only the underlying torus that is factorizable, whereas the orbifold is not $(T^2/\Z_{\N_1})\times(T^2/\Z_{\N_2})\times(T^2/\Z_{\N_3})$.

The string states on the orbifold satisfy closed string boundary conditions that incorporate a non-trivial local monodromy:
\be
X^i (e^{2\pi\imath}z,e^{-2\pi\imath}\bar z) = (\theta^k X)^i(z,\bar z) + \lambda^i  \equiv (g X)^i(z,\bar z) \,,\label{E:Xbc}
\ee
with $X^i$, $i=1,2,3$, the complex coordinates of the string in the target orbifold space, and $z,\bar z$ the complex worldsheet coordinates.
When $k=0$ we call the states untwisted, otherwise they are twisted, in the $k$-th twisted sector.  A twisted state is associated, via Eq. (\ref{E:Xbc}), not just to a single space group element $g$, but with an entire conjugacy class of the space group, where $\left\{hgh^{-1}\, | \, h\in S\right\}$ is the conjugacy class of $g$ \cite{Dixon}.  

For prime orbifolds, the conjugacy classes are in one-to-one correspondence with the fixed points of $P$ and are given by $\left\{(\theta^k, \lambda + (1-\theta^k)\Lambda)\right\}$, with the translation part running over some coset of the torus lattice $\Lambda$.  For non-prime orbifolds, it may happen that a point $f$ fixed under $\theta^k$ is not fixed under $\theta$.  We can then consider its orbit ${\mathcal O}_f$ as $\left\{\theta^r f; 0 \leq r < l\right\}$, for $l$ the smallest integer such that $\theta^l f = f + \lambda$.  
In other words, two points fixed under $\theta^k$ that are different on the torus may be connected by $\theta$, in which case both correspond to the same conjugacy class 
and are equivalent in the orbifold.  The conjugacy classes are then given by:
\be
\bigcup_{r=0}^{l-1}\left\{ (\theta^k, \theta^r \lambda + (1-\theta^k)\Lambda)\right\} \,,
\ee
with higher twisted sectors generically corresponding to a union of cosets.  It follows that for non-prime orbifolds, the physical states in the $k$-th twisted sector are in general linear combinations of states located at several $\theta^k$ fixed points \cite{Kobayashi:1991rp,physical,CGM}. Labelling states by their associated fixed points we construct the physical states $|\psi\rangle$ as
\be
|\psi\rangle = |f\rangle + e^{-2\pi\imath \gamma}|\theta f \rangle + \dots + e^{-2\pi\imath (l-1)\gamma}|\theta^{l-1} f \rangle \,, \label{psi}
\ee
with the possible $\gamma$-phases given by $\gamma= p/l$, $p = 0,1,\dots,l-1$.  The geometric part of the string state is thus constructed to be an eigenstate of the $\theta$ twist, with eigenvalue given by $e^{2\pi\imath \gamma}$.

So far we have presented the action of the $\Z_\N$ point group on the bosonic coordinate fields $X^i$, but modular invariance requires the twist to act non-trivially on the other components of the heterotic string.  In particular, excluding Wilson lines, the twist is embedded into the gauge group with the gauge degrees of freedom transforming as $X^I \rightarrow X^I + 2\pi V^I$, $I=1,\dots, 16$, where $V^I$ is called the gauge shift vector.

The physical string states must be invariant under the total $\Z_\N$ transformation.  We will show this in more detail below, in terms of the corresponding vertex operators.

\subsection{Vertex operators and correlation functions}

The physical states in the string Hilbert space correspond to fields or operators in the orbifold CFT.  To compute the low energy effective field theory, we are interested in the vertex operators describing the emission of massless fields, in the limit of zero 4D momentum.  For a twisted bosonic field, this is given by\footnote{We omit normalization and cocycle factors, as they are not important for our purposes.}:

\be
V_{-1} = e^{-\phi} \prod_{i=1}^3 (\partial X^i)^{{\cal N}_L^i} \,\,
(\partial \bar X^i)^{\bar{\cal N}_L^i} \,\, e^{\imath q_{sh}^{m} H^{m}}
\, e^{\imath p_{sh}^I X^I} \,
\sigma^{i}_{\left(k,\psi\right)} \,, \label{V_-1}
\ee
whilst for a twisted fermionic field it is

\be
V_{-1/2} = e^{-\phi/2} \prod_{i=1}^3 (\partial X^i)^{{\cal N}_L^i}
\,\, (\partial \bar X^i)^{\bar{\cal N}_L^i} \,\, e^{\imath q_{sh}^{(f)\,m}
H^m} \, e^{\imath p_{sh}^I X^I} \,
\sigma^{i}_{\left(k,\psi\right)}  \,.
\ee
Each has a number of contributions, corresponding to the various degrees of freedom carried by the heterotic string.  
The right-moving worldsheet fermions have been bosonized to the fields $H^m(\bar z)$, $m=1,\dots,5$, and they carry an H-momentum given by $q_{sh}=q+kv$ for the twisted spacetime bosonic states, with $q$ a weight on the vector lattice of $SO(10)$, and $k v$ the shift induced by the orbifold twist.  The H-momentum for the spinor representation is 
written as $q_{sh}^{(f)}$, and  is related to that in the
vector representation by:
\be
q_{sh} =  q_{sh}^{(f)} + (1,1,1,1,1)/2 \,. \label{qsh}
\ee  
The left-moving worldsheet bosonic gauge fields correspond to $X^I(z)$, $I=1,\dots,16$, which carry a similarly shifted gauge momentum, $p_{sh} = p + kV$, where $p$ is a vector in the $E_8 \times E_8$ lattice.  Massless states may be excited via some number, $\calN_L^i, \bar\calN_L^i$, of left-moving holomorphic and anti-holomorphic bosonic oscillators.  We have also introduced a scalar field, the superconformal ghost $\phi$, with the subscript on $V$ indicating the superconformal ghost charge.  

The twist fields $\sigma_{(k,\psi)}$ serve to implement the non-trivial monodromy (\ref{E:Xbc}) observed for the bosonic string coordinates $X^i(z,\bar z)$.    For the prime orbifolds there is a one-to-one correspondence between the twist fields and fixed points, but for non-prime orbifolds it is helpful to construct them as \cite{twistfields}:
\be
\sigma_{(k,\psi)} = \sum_{r=0}^{l-1} e^{-2\pi\imath r \gamma}\sigma_{(k,\theta^r f)} \,,
\ee
with $\sigma_{(k,\theta^r f)}$ being viewed as auxiliary twist fields. Note that, as far as quantum properties go, the auxiliary twist fields $\sigma_{(k,\theta^r f)}$ with various $\theta^r f$ are indistinguishable.  
Also, untwisted fields, which have $k=0$, are included in our discussion by taking $\sigma^i_{(k,f)} \to 1$.  

The quantum numbers of the physical massless states should fulfill the masslessness conditions and orbifold projection conditions.   Under the twist, the complete state associated with the space group element $g=(\theta^k,\lambda)$ acquires a phase:
\be
\Delta(k,e^{2\pi\imath\gamma}) = e^{2\pi\imath \left[(\calN_L-\bar\calN_L).v + p_{sh}.V - q_{sh}.v + \gamma - \frac{k}{2} \left(V^2-v^2\right)\right]} \,, \label{GSO}
\ee
so $\Delta(k,e^{2\pi\imath\gamma})=1$ for physical states.

With the vertex operators at hand, we are ready to consider the correlation functions.  We are interested in string tree-level L-point correlation functions of the kind $\langle V_F V_F V_B \dots V_B \rangle$, as this is enough to identify allowed terms in the 
holomorphic superpotential of the low energy effective field theory.  To cancel the background ghost-charge, which is 2 on the worldsheet sphere, we have to introduce $\L-3$ picture-changing operators into the correlation function, taking $\langle V_{-1/2} V_{-1/2} V_{-1} V_0 \dots V_0 \rangle$ with

\bea
\hskip-0.8cm V_{0} &=&\sum_{j=1}^3\! \left(e^{\imath q_{0}^{j\,m} H^m} \,
\bar\partial X^j \!+ \!e^{-\imath q_0^{j \,m} H^m} \,\bar\partial \bar X^j \!\right) \cr
&& \qquad \qquad \times  \prod_{i=1}^3
\!(\partial X^i)^{{\cal N}_L^i} \,
(\partial \bar X^i)^{\bar{\cal N}_L^i } \,\, e^{\imath q_{sh}^m H^m} 
e^{\imath p_{sh}^I X^I} \!\sigma^{i}_{\left(k,\psi\right)}.
\label{V0} 
\eea
Thus, for higher order couplings, with $\L>3$, additional H-momenta $q_0^{1} = (0,0,1,0,0)$, $q_0^{2}= (0,0,0,1,0)$ and $q_0^{3} = 
(0,0,0,0,1)$ need to be introduced, as well as right-moving oscillators, which we  count by $\calN_R^i, \bar\calN_R^i$.

\subsection{String selection rules}
\label{S:calF}

We now review how string coupling selection rules can be derived from the string correlation functions \cite{Dixon, HV, KPRZ}.  In particular, in this section we recall the space group selection rule, gauge invariance, H-momentum conservation and Rule 5, referring the reader to \cite{KPRZ} for more details.  In the following sections we will turn to the main topic of 
the paper; R-charge conservation and a new stringy rule, as well as Rule 4, in factorizable and non-factorizable orbifolds.

Since the orbifold CFT is free, the L-point correlation functions factorize into several parts, each giving rise to a certain condition for non-vanishing couplings.  The twist fields lead to the so-called space group selection rule, which takes the form ($\alpha = 1, \dots, \L$):
\be
\prod_{\alpha=1}^\L \left[ g_{\alpha} \right] = (\mathbbm{1},0) \,,
\ee
where $ \left[ g_{\alpha} \right]$ represents some element of the conjugacy
class of the space group element $g_\alpha$.  The space group selection
rule includes the point group selection rule, $\sum_{\alpha=1}^\L
k_\alpha = 0 \mod \N$ for the $\Z_\N$ orbifold.   In terms of the explicit space group elements, it can be written as\footnote{In a slight abuse of notation, throughout the text we  use $\theta$ to denote both the generator of $P$ and the Coxeter element of the lattice Lie algebra, which gives the action of the point group on the basis vectors of the lattice. The meaning should however be clear from the context. }
\bea
\!\!\!\!\! &&\!\!\!\!\!(1-\theta^{k_\L})(\theta^{r_{\L}}f_\L + \tau_\L) + \theta^{k_\L}(1-\theta^{k_{\L-1}})({\theta^{r_{\L-1}}}f_{\L-1} + \tau_{\L-1}) 
+ \ldots \nn \\
&&\hskip4cm \ldots + \theta^{k_{\L}+ k_{\L-1}+ \dots + k_2}(1-\theta^{k_1})({\theta^{r_{1}}}f_1 + \tau_1) = 0 \,, \label{SGSR}
\eea
for some integers $0\leq r_\alpha<\N$ and some $\tau_{\alpha} \in \Lambda$. Thus we see 
that the space group selection rule restricts the combinations of fixed points 
that can enter a coupling.

Next we have the momentum conservation conditions.  The conservation of gauge momentum:
\be
\sum_\alpha p_{sh} = 0,
\ee
leads to gauge invariance.  Instead, conservation of H-momentum implies that  all terms in the correlation function vanish except for those satisfying:
\be
\calN_R^i = 0, \qquad \bar\calN_R^i = \sum_\alpha q_{sh \,\alpha}^i - 1 \,. \label{Hmom}
\ee
Note that for higher order couplings, H-momentum conservation is automatically satisfied after imposing the point group selection rule, but must be considered to fix the number of right-moving oscillators in the correlation function.  
We will comment on H-momentum conservation for three-point couplings below.

After applying the momentum conservation, the non-trivial part of the general correlation function takes the form \cite{Rule4, cvetic, KPRZ}:
\be
{\mathcal F} = \prod_{i=1}^3 \langle (\partial X^i)^{{\cal N}_L^i}
\,\, (\partial \bar X^i)^{\bar{\cal N}_L^i}  \,\, (\bar\partial
\bar X^i)^{\bar{\cal N}_R^i}  
\sigma_{\left(k_1,\psi_1\right)}^i \cdots
\sigma_{\left(k_\L,\psi_\L\right)}^i 
\rangle \, , \label{beforesplit}
\ee
where we have factorized it into 2D components, and from now on $\calN_L^i, \bar\calN_L^i$, like $\bar\calN_R^i$, refer to the total number of oscillators appearing in the correlation function.   Notice moreover that for the higher twisted sectors, we can write the correlation function for physical states as a sum of several auxiliary correlation functions, each involving the auxiliary twist fields, weighted by the $\gamma$-phases \cite{Dixon, erler}:

\bea
{\mathcal F} &=& \sum_{r_1=0}^{l_1}\cdots\sum_{r_\L=0}^{l_\L} e^{-2\pi\imath r_1\gamma_1 - \dots -2\pi\imath r_\L\gamma_\L} \times\prod_{i=1}^3  {\mathcal F}_{aux}^i\,,  \label{auxdecomp}
\eea
where
\bea
{\mathcal F}_{aux}^i = \langle (\partial X^i)^{{\cal N}_L^i}
\,\, (\partial \bar X^i)^{\bar{\cal N}_L^i}  \,\, (\bar\partial
\bar X^i)^{\bar{\cal N}_R^i}  \sigma^i_{(k_1,\theta^{r_1}f_1)} \cdots \sigma^i_{(k_\L,\theta^{r_\L}f_\L)}
\rangle \, . 
\eea

To make further progress, it is helpful to split the bosonic coordinates into the solutions to their classical equations of motion, $\partial\bar\partial X^i_{cl} = 0$, and their quantum fluctuations:
\be
X^i(z,\bar z) = X^i_{cl}(z,\bar z) + X^i_{qu}(z,\bar z) \,.
\ee
The auxiliary correlation functions similarly split as:
\bea
&&{\mathcal F}^i_{aux} = \sum_{s=0}^{ {\cal N}_L^i} {\scriptsize\begin{pmatrix} {\cal N}_L^i \\ s \end{pmatrix} }  \sum_{t=0}^{{\bar{\cal N}}_L^i}  {\scriptsize\begin{pmatrix} {\bar{\cal N}}_L^i \\ t \end{pmatrix}}\sum_{u=0}^{\bar{\cal N}_R^i} {\scriptsize\begin{pmatrix} \bar{\cal N}_R^i \\ u \end{pmatrix}} 
\sum_{X^i_{cl}} e^{-S^i_{cl}} (\partial X^i_{cl})^{{\cal 
N}_L^i-s}  \,\,(\partial \bar X^i_{cl})^{\bar{\cal N}_L^i-t}  \,\, (\bar\partial \bar X^i_{cl})^{\bar{\cal N}_R^i-u}  \,\, \nonumber \\
&& \qquad \qquad \quad \times \; \int {\mathcal D}X^i_{qu} e^{-S^i_{qu}} (\partial X^i_{qu})^s  \,\,(\partial \bar
X^i_{qu})^t\,\, (\bar\partial \bar X^i_{qu})^u 
\sigma^i_{\left(k_1,\theta^{r_1}f_1\right)} \cdots
\sigma^i_{\left(k_\L,\theta^{r_\L}f_\L\right)}\,,  \label{Fqu+Fcl} 
\eea
where $\scriptsize{\begin{pmatrix} {\cal N}_L^i \\ s \end{pmatrix}}$ and so on are the binomial coefficients and the classical action is given by 

\be
S^i_{cl}= \frac{1}{8\pi} \int d^2z \left(|\partial
X^i_{cl}|^2 + |\partial \bar X^i_{cl}|^2 \right).
\ee 
Moreover, for the quantum part of the correlation functions to be non-vanishing, the number of holomorphic and anti-holomorphic indices have to match \cite{HV}, $s=t+u$.  Thus, in fact:
\bea
&&{\mathcal F}^i_{aux} = \sum_{t=0}^{\bar{\cal N}_L^i} \sum_{u=0}^{\bar{\cal N}_R^i}{\scriptsize \begin{pmatrix} {\cal N}_L^i \\ t+u \end{pmatrix}}    {\scriptsize\begin{pmatrix} {\bar{\cal N}}_L^i \\ t \end{pmatrix}} {\scriptsize\begin{pmatrix} {\bar{\cal N}}_R^i \\ u \end{pmatrix}}
\sum_{X^i_{cl}} e^{-S^i_{cl}} (\partial X^i_{cl})^{{\cal 
N}_L^i-t-u}  \,\,   (\partial \bar X^i_{cl})^{\bar{\cal N}_L^i-t}  \,\, (\bar\partial \bar X^i_{cl})^{\bar{\cal N}_R^i-u}  \label{Fqu+Fcl2} \nonumber\\
&& \qquad  \times \int {\mathcal D}X^i_{qu} e^{-S^i_{qu}}  (\partial X^i_{qu})^{t+u} \,\,(\partial \bar X^i_{qu})^{t} \,\,(\bar \partial \bar
X^i_{qu})^{u}
\sigma^i_{\left(k_1,\theta^{r_1}f_1\right)} \cdots
\sigma^i_{\left(k_\L,\theta^{r_\L}f_\L\right)}  \,, \label{E:s=t+u}
\eea
where the non-zero contributions in the sums over $t$ and $u$ satisfy $t+u \leq {\cal N}_L^i$.

The classical solutions represent worldsheet instantons stretching between the fixed points involved in the auxiliary couplings, and can be written as \cite{Dixon,ADGN,erler}:
\bea
\partial X^i_{cl}(z) &=& \sum_{l=1}^{\L-\M^i-1} a_{l}^i \,  h_{l}^i(z),
\label{dXcl}\\ 
\bar\partial X^i_{cl}(\bar z) &=&  \sum_{l'=1}^{\M^i-1} b_{l'}^i  
\, 
\bar {h'}_{l'}^{i}(\bar z),  \label{bardXcl}
\eea
(no summation over $i$), plus their complex conjugates: 
\bea
\bar\partial \bar X^i_{cl} &=& \left( \partial X^i_{cl} \right)^*, \cr
\bar\partial X^i_{cl} &=& \left( \partial \bar X^i_{cl} \right)^* \, .
\eea
Here,  the  integers $\M^i$ are given by $\M^i=\sum_{\alpha=1}^\L
\k_\alpha^i$, and we have defined $\k_\alpha^i = k_\alpha  \, v^i \mod
1$, such that $0 < \k_\alpha^i \leq 1$  in (\ref{dXcl}),
and $0 \leq \k_\alpha^i < 1$ in (\ref{bardXcl}).  Any untwisted strings present in the coupling do not alter these instanton solutions.

The basis functions $h_l^i(z)$, $\bar {h'}_{l'}^i(\bar z)$ are determined by the local monodromy conditions (\ref{E:Xbc}), and the requirement that the classical action converges \cite{ADGN}.
It can happen that the only way to satisfy these conditions is with a vanishing solution $\partial X^i_{cl} = 0$ or $\partial \bar X^i_{cl} = 0$.  
In particular, at least three twisted strings in a given plane are required for a non-trivial worldsheet instanton to exist in that plane.
This implies Rule 5 or the {\it forbidden instanton rule}, which imposes the following conditions on the oscillator numbers \cite{KPRZ}:
\bea
&&\textrm{holomorphic instantons forbidden:  }  {\cal N}_L^i \leq \bar{\cal N}_L^i+\bar{\cal N}_R^i \nonumber\\
&&\textrm{anti-holomorphic instantons forbidden:  } {\cal N}_L^i \geq \bar{\cal N}_L^i \nonumber \\
&&\textrm{no instantons allowed:  }  {\cal N}_L^i = \bar{\cal N}_L^i + \bar{\cal N}_R^i \,,
\eea
and simplifies further the expression (\ref{E:s=t+u}) for the correlation function.  In detail, for holomorphic instantons forbidden, the auxiliary correlation functions in the $i$-th plane take the form:
\bea
&&{\mathcal F}^i_{aux} = {\scriptsize\begin{pmatrix} {\bar{\cal N}}_L^i \\ {\cal N}_L^i- {\bar{\cal N}}_R^i\end{pmatrix}} 
\sum_{X^i_{cl}} e^{-S^i_{cl}} (\partial \bar X^i_{cl})^{{\bar{\cal N}}_L^i-{\cal N}_L^i+{\bar{\cal N}}_R^i}  \label{E:nohol}\\
&& \qquad  \quad \times \int {\mathcal D}X^i_{qu} e^{-S^i_{qu}}  (\partial X^i_{qu})^{{\cal N}_L^i } \,\,(\partial \bar X^i_{qu})^{{\cal N}_L^i -{\bar{\cal N}}_R^i} \,\,(\bar \partial \bar
X^i_{qu})^{\bar{\cal N}_R^i}
\sigma^i_{\left(k_1,\theta^{r_1}f_1\right)} \cdots
\sigma^i_{\left(k_\L,\theta^{r_\L}f_\L\right)}  \,, \nonumber
\eea
for anti-holomorphic instantons forbidden, we have:
\bea
&&{\mathcal F}^i_{aux} = \sum_{t=0}^{{\rm min}({\bar{\cal N}}_R^i,{\cal N}_L^i-{\bar{\cal N}}_L^i)} {\scriptsize\begin{pmatrix} {\cal N}_L^i \\ {\bar{\cal N}}_L^i+t\end{pmatrix}} {\scriptsize\begin{pmatrix} {\bar{\cal N}}_R^i \\ t\end{pmatrix}} 
\sum_{X^i_{cl}} e^{-S^i_{cl}} (\partial X^i_{cl})^{{\cal N}_L^i-{\bar{\cal N}}_L^i-t} (\bar \partial \bar X^i_{cl})^{{\bar{\cal N}}_R^i-t}  \label{E:noantihol}\\
&& \qquad  \quad \times \int {\mathcal D}X^i_{qu} e^{-S^i_{qu}}  (\partial X^i_{qu})^{{\bar{\cal N}}_L^i+t} \,\,(\partial \bar X^i_{qu})^{{\bar{\cal N}}_L^i} \,\,(\bar \partial \bar
X^i_{qu})^{t}
\sigma^i_{\left(k_1,\theta^{r_1}f_1\right)} \cdots
\sigma^i_{\left(k_\L,\theta^{r_\L}f_\L\right)}  \,, \nonumber
\eea
and for no instantons allowed:
\be
{\mathcal F}^i_{aux}= \int {\mathcal D}X^i_{qu} e^{-S^i_{qu}}  (\partial X^i_{qu})^{{\cal N}_L^i} \,\,(\partial \bar X^i_{qu})^{{\bar{\cal N}}_L^i} \,\,(\bar \partial \bar
X^i_{qu})^{{\bar{\cal N}}_R^i}
\sigma^i_{\left(k_1,\theta^{r_1}f_1\right)} \cdots
\sigma^i_{\left(k_\L,\theta^{r_\L}f_\L\right)}  \,. \label{E:noinst}
\ee

To obtain further selection rules, the coefficients $a_l, b_{l'}$ in the worldsheet instanton solutions will be important.  These are determined by the global monodromy conditions (the quantum part instead feels only the local monodromy) \cite{Dixon, dXcl,erler}:     

\begin{eqnarray}
\int_{\Gamma_p} dz \partial X^i_{cl} + \int_{\Gamma_p} d\bar z
\bar\partial X^i_{cl} = \nu_p^i , \label{globmon} \cr 
\int_{\Gamma_p} dz \partial \bar X^i_{cl} + \int_{\Gamma_p} d\bar z
\bar\partial \bar 
X^i_{cl} = \bar{\nu}_p^i \,,
\end{eqnarray}
where $\Gamma_p$ represent all possible net zero-twist closed loops enclosing the twist fields.   There will be L-2 independent such loops, which we can take to be those encircling the fixed point $f_p$ clockwise $n_p$ times followed by the fixed point $f_{p+1}$ counterclockwise $m_p$ times, where $n_p k_p = m_p k_{p+1} \mod \N$, with $n_p, m_p$ the smallest integers satisfying this property, and $p=1,\dots,\L-2$.  The vectors $\nu_p^i$  are then elements of the corresponding cosets of the torus lattice:
\be
\nu_p =
(1-\theta^{n_p \,k_p})(f_{p+1}-f_p+\lambda)\,, \qquad \lambda \in \Lambda \,.
\label{cosets} 
\ee
Note that the $(\L-1)$-th loop is not independent from the others, as the sum of all $\L-1$ loops can be pulled around the worldsheet sphere and shrunk to zero, giving the space group selection rule.  However, this provides a consistency constraint that may further restrict the coset vectors $\nu_p^i$.  For example, for the 3-point couplings it turns out that the coset vectors are restricted to \cite{erler}:
\be
\nu_1 = \left(1-\theta^{n_1 \,k_1}\right) \left(f_2 - f_1  - \tau_2 + \tau_1 + (1-\theta^{k_1+k_2}){(1-\theta^{{\rm gcd}(k_1,k_2)})^{-1}} \lambda \right)\,,  \quad \lambda \in \Lambda\,,
\label{3ptcoset}
\ee
where $\tau_{1,2}$ are the torus lattice vectors that appear in the space group selection rule (\ref{SGSR}) and 
{\rm gcd} stands for greatest common divisor.  Finally, we can solve the global monodromy conditions for the coefficients
$a_l^i, b_{l'}^i$.  Defining the period matrices as:
\bea
W_{\,\,\,p}^{i\,l} &=& \int_{\Gamma_p} dz h_l^i(z), \qquad l = 1,
\dots, \L-\M^i-1, \cr 
W_{\,\,\,p}^{i\, (\L-\M^i-1+l')} &=& \int_{\Gamma_p} d\bar z \bar
{h'}_{l'}^i(\bar z), \qquad 
l' = 1, \dots, \M^i-1  \,,
\eea
it follows that $a_l^i, b_{l'}^i$ are particular linear combinations of the coset
vectors $\left\{\nu^i_p \right\}$:
\bea
a_l^i &=& \nu_p^i (W^{-1})^{i\,p}_{\,\,\,l} \,, \cr
b_{l'}^i &=&{\nu}_p^i (W^{-1})^{i\,p}_{\,\,\,L-M^i-1+l'} \,.
\label{a,b} 
\eea
Notice that the fixed point dependence lies only in the coset vectors $\nu_p^i$, as the period matrices depend only on the twisted sectors involved.

We are now ready to derive more selection rules, which arise due to the symmetries amongst the worldsheet instanton solutions.  
To this purpose, we use Eqs. (\ref{E:s=t+u}) and (\ref{E:noantihol}) to show that, when holomorphic instantons are allowed in the $i$-th plane, the auxiliary correlation functions can always be written in the form:
\be
\calF^i_{aux} = \sum_{X_{cl}^i} e^{-S_{cl}^i}(\partial X^i_{cl})^{{\cal 
N}_L^i-\bar{\cal  N}_L^i-\bar{\cal N}_R^i} {\mathit f}(|\partial X_{cl}^i|^2,\partial X_{cl}^i \partial \bar X_{cl}^i) \, \langle\sigma^i_{(k_1,\theta^{r_1}f_1)}\dots\sigma^i_{(k_\L,\theta^{r_\L}f_\L)}\rangle \,, \label{E:Faux}
\ee
where the explicit expression for ${\mathit f}(|\partial X_{cl}|^2, \partial X_{cl}^i \partial\bar X_{cl}^i)$ follows from  Eqs. (\ref{E:s=t+u}) and (\ref{E:noantihol}), and depends on the classical solutions as written, and we use the shorthand $\langle\sigma^i_{\theta^{r_1}f_1}\dots\sigma^i_{\theta^{r_\L}f_\L}\rangle$ for the quantum part of the correlation function.   Similarly, when holomorphic instantons are forbidden in the $i$-th plane, it follows from Eqs. (\ref{E:nohol}) and (\ref{E:noinst}) that we can write the auxiliary correlation functions as:
\be
\calF^i_{aux} = \sum_{X_{cl}^i} e^{-S_{cl}^i}(\partial \bar X^i_{cl})^{-{\cal 
N}_L^i+\bar{\cal  N}_L^i+\bar{\cal N}_R^i}  \, \langle\sigma^i_{(k_1,\theta^{r_1}f_1)}\dots\sigma^i_{(k_\L,\theta^{r_\L}f_\L)}\rangle \,.\label{E:Faux2}
\ee
It is important to observe that the classical solutions depend on the fixed points to which the participating auxiliary twist fields are associated.  In detail, since the basis functions $h_l^i, h_{l'}^{i'}$ are determined by the local monodromy, they depend only on the twisted sectors involved in the coupling, and not the fixed points.  Instead, the coefficients $a^i_l, b^{i}_{l'}$ are determined by the global monodromy, and so do depend on the fixed point positions.  If the fixed points involved in two couplings are related by the orbifold twist, then so will be the coefficients $a^i_l, b^i_{l'}$ and hence also the classical solutions. Notice also that given two classical solutions, $\partial X^i_{cl\,1}$, $\partial X^{i}_{cl\,2}$ related by $\theta$, then $|\partial X^i_{cl\,1}|^2 = |\partial X^{i}_{cl\,2}|^2$ and $\partial X^i_{cl\,1} \partial\bar X^i_{cl\,1} = \partial X^{i}_{cl\,2} \partial\bar X^{i}_{cl\,2}$.

\section{R-charge Conservation, the $\gamma$-Phase and a New Stringy Rule}
Our purpose is now to use what we have learned to study the origin and structure of the R-charge conservation laws.  R-charge conservation in orbifold compactifications has been understood from two related perspectives \cite{Rule4, Rcharge&miracle, Rcharge}.  Since Lorentz symmetries distinguish between bosons and fermions, one expects any discrete Lorentz symmetries in the internal space that survive the orbifold compactification to lead to a discrete R-charge conservation law in the four dimensional effective field theory.  On the more technical side, the R-charge conservation law has been derived from the general structure of the string correlation functions, as a consequence of the H-momentum conservation condition and the
plane by plane twist invariance that is observed in factorizable prime orbifolds \cite{KPRZ}.

We now refine these arguments, considering carefully the various classes of $\Z_\N$ orbifolds, prime, non-prime, factorizable, non-factorizable.  In doing so, we find that
the relation between Lorentz symmetries in the orbifold geometry and R-symmetries in the low energy effective field theory can be more subtle than previously assumed. Also,  
the R-charges of higher twisted states in non-prime orbifolds have a non-trivial contribution from their $\gamma$-phase.  Moreover, 
for certain orbifolds we find an additional selection rule whose string origin is similar to that of the R-charge conservation law.  This rule, which we may call Rule 6, depends however on the relative properties of the states in a coupling, and hence does not have a simple field theoretical interpretation.

\subsection{Orbifold automorphisms}
\label{S:lorentz}

To begin with, we are interested in the symmetries of the internal space of the string compactification, that is the orbifold $T^6/\Z_\N$.  Thus we are looking for the subgroup of the torus lattice automorphism group that respects the point group and moreover leaves the conjugacy classes of the fixed points invariant.  We focus on the Lorentz transformations that are discrete rotations. 
In the appendix, we discuss more generally the symmetries of the orbifold geometry.

When the underlying torus lattice is factorizable, the action of the orbifold twist $\theta$ can be decomposed into discrete rotations acting plane by plane, $(\theta_1,\theta_2,\theta_3)$, and the fixed points can be decomposed as a direct product of fixed points in the three planes, $f^{(k)}=g_1 \otimes g_2 \otimes g_3$, with $(\theta_i)^k g_i = g_i$ (up to lattice identifications) for $f^{(k)}$ in the $k$-th twisted sector.  We can then immediately identify discrete rotational symmetries.  First, notice that for a plane with prime order twist, $\theta_i g_i = g_i$ for all the twisted sectors.  Then, there are three different cases:
\begin{enumerate}[(i)]

\item \label{item1} {\it All planes have prime order twists.}  In this case, all the fixed points are fixed under the orbifold twist plane by plane:
\be
\theta_i f^{(k)} = f^{(k)}, \quad i=1,2,3, \quad \textrm{for all } f^{(k)} \,. \label{allprime}
\ee
So we have the discrete symmetries generated by $\theta_1, \theta_2$ and $\theta_3$.  An example of this case is $T^6/\Z_3$.

\item \label{item2}{\it Only one plane is non-prime.}  Here, all fixed points are fixed under the prime plane rotations, say $\theta_2, \theta_3$.  Moreover, considering the non-prime rotation, say $\theta_1$, we have:
\bea
\theta_1 f^{(k)} &=& \theta_1 g_1 \otimes g_2 \otimes g_3 \nonumber \\
&=& \theta_1 g_1 \otimes \theta_2 g_2 \otimes \theta_3 g_3 \nonumber\\
&=& \theta f^{(k)} \nonumber \\
&\simeq& f^{(k)}\,. 
\eea
where the last $\simeq$ indicates equivalence of the fixed points up to the conjugacy class.  So we have again the symmetries, $\theta_1, \theta_2$ and $\theta_3$.  Here, an example is $T^6/\Z_{6-II}$.

\item \label{item3}{\it Two planes are non-prime.}  Again, all the fixed points are fixed under the prime plane rotation, say $\theta_3$.  Moreover, they are invariant under the combined action of the non-prime rotations, say $\theta_1 \theta_2$ since:
\bea
\theta_1 \theta_2 f^{(k)} &=& \theta_1 g_1 \otimes \theta_2 g_2 \otimes g_3 \nonumber \\
&=& \theta_1 g_1 \otimes \theta_2 g_2 \otimes \theta_3 g_3 \nonumber\\
&=& \theta f^{(k)} \nonumber \\
&\simeq& f^{(k)}\,.
\eea 
In this case, the symmetries are generated by $\theta_1\theta_2$ and $\theta_3$.  An example is $T^6/\Z_{6-I}$.
\end{enumerate}

There is one more case to be considered.  The $\Z_4$ orbifold, with twist vector \linebreak\hbox{$v=\frac14(1,1,-2)$}, has only two kinds of twisted sectors, $\theta, \theta^2$ (plus their conjugates).  Therefore, all the fixed points are fixed under $\theta^2$, and this holds plane by plane for the factorizable orbifold:
\be
(\theta)^2 f^{(k)} =
(\theta_1)^2 g_1 \otimes (\theta_2)^2 g_2 \otimes (\theta_3)^2 g_3 
= g_1 \otimes g_2 \otimes g_3 = f^{(k)}\Rightarrow \nonumber
\ee
\be
(\theta_1)^2 f^{(k)} = f^{(k)} \,, \quad (\theta_2)^2 f^{(k)} = f^{(k)} \, . \label{Rule6factorizable}
\ee
Thus there will be an additional symmetry generated by $(\theta_1)^2$ or $(\theta_2)^2$ (these are not independent).

For orbifolds whose underlying torus lattice and orbifold twist are partially factorizable on one complex plane, a similar analysis can be made.  For the non-factorizable orbifolds, the twist does not act plane by plane, and we cannot decompose the fixed points as a direct product of fixed points in the three planes.  We may still ask what are the discrete rotational symmetries of the orbifold geometry, that is which elements of the Cartan subalgebra of $SO(6)$ leave the torus lattice and fixed point conjugacy classes invariant.  We have performed a computer scan for such orbifold automorphisms in several explicit examples,
with results given in Table \ref{T:nonfact}. In most cases, the only symmetry is the point group itself, but notice that for the $\Z_4$ and $\Z_{8-I}$ orbifolds, the fixed points are such that they enjoy an additional $\Z_2$ symmetry generated by $(\theta_1)^2$. A similar computer scan for the factorizable orbifolds revealed that the list of possible symmetries given in~(\ref{item1})-(\ref{item3}), is exhaustive.

\TABLE{
\centering 
\renewcommand{\arraystretch}{1.4}
\scriptsize{
\begin{tabular}{|c||c|c|c|} 
\hline
& Lattice & Twist & {Orbifold Automorphisms} \\
\hline\hline
$\mathbbm{Z}_4$ & SU(4)$\otimes$SU(4) & $\frac{1}{4}(1,1,-2)$ & $\theta$, $(\theta_1)^2$\\
\hline
$\mathbbm{Z}_{6-\mathrm{II}}$ & SU(6)$\otimes$SU(2) & $\frac{1}{6}(1,2,-3)$ & $\theta$ \\
\hline
$\mathbbm{Z}_{7}$ & SU(7) & $\frac{1}{7}(1,2,-3)$ & $\theta$ \\
\hline
$\mathbbm{Z}_{8-\mathrm{I}}$ & SO(5)$\otimes$SO(9) & $\frac{1}{8}(2,1,-3)$ & $\theta$, $(\theta_1)^2$ \\
\hline
$\mathbbm{Z}_{8-\mathrm{II}}$ & SO(8)$\otimes$SO(4) & $\frac{1}{8}(1,3,-4)$ & $\theta$, $\theta_3$ \\
\hline
$\mathbbm{Z}_{12-\mathrm{I}}$ & SU(3)$\otimes F_4$ & $\frac{1}{12}(4,1,-5)$ & $\theta$, $\theta_1$\\
\hline
$\mathbbm{Z}_{12-\mathrm{II}}$ & $F_4\otimes $SO(4) & $\frac{1}{12}(1,5,-6)$ & $\theta$, $\theta_3$\\
\hline
\end{tabular}}
\caption{Orbifold automorphisms for some non-factorizable or partially factorizable orbifolds
built on Lie lattices, counting independent discrete rotational symmetries that preserve the conjugacy classes, and labeling them with their generators.
We refer to \cite{Kobayashi:1991rp,CGM} for details of the torus lattice, orbifold twist and fixed points.}
\label{T:nonfact}
}

In what follows, we show how these symmetries in the orbifold geometry affect the string couplings.  
For smooth, field theoretical compactifications, we know that Lorentz symmetries of the compact manifold give rise to R-symmetries in the low energy effective field theory, and that the 4D R-charges descend from the 10D Lorentz representations.  This intuition has always been applied also to orbifold string compactifications, but has been worked out explicitly only in some special cases.  In particular, we should check how the twisted sectors, which emerge only after the orbifold compactification and are localized in the orbifold geometry, transform under the effective R-symmetry.  For example, referring to the expression for the vertex operator Eq. (\ref{V_-1}), if we were to suppose naively that the only components of the string that transform under the 6D Lorentz transformation are $X^i$ and $H^i$, with $\partial X^i \rightarrow e^{2\pi\imath v_i} \partial X^i$, $\partial \bar X^i \rightarrow e^{-2\pi\imath v_i} \partial \bar X^i$ and $H^i \rightarrow H^i - 2\pi v^i$ under the Lorentz rotation $\theta_i$, we would infer that the picture-independent R-charges,  \hbox{$R^i = q_{sh}^i -\calN_L^i+\bar\calN_L^i$}, should be conserved \cite{Rcharge&miracle}.  However, we would not recover in general the required transformation under the full orbifold twist\footnote{The same can be said for untwisted charged matter, as the orbifold twist acting in all planes together $\theta_1\theta_2\theta_3=\theta$, should be embedded into the gauge sector.} $\theta_1\theta_2\theta_3=\theta$.  To identify the R-symmetry in the effective field theory, including the R-charges carried by twisted fields, we have to look at the explicit expressions for the correlation functions.

\subsection{A 2D illustrative example}
\label{S:2d}
It will be useful to have a simple 2D example in mind, before considering the general 6D case.  As the prime orbifolds follow straightforwardly, we present a non-prime example, $T^2/\Z_6$ on a 
$G_2$ lattice (see Fig.~\ref{2DEjemplo}).  We focus for simplicity on the 3-point couplings, where the explicit form of the worldsheet instanton solutions mediating the couplings is known completely.

We describe the $G_2$ lattice with basis vectors, $e_1, e_2$.  The $\Z_6$ twist acts in the lattice basis as
\be
\theta e_1 = -e_1 -e_2, \qquad \theta e_2 = 3 e_1 + 2 e_2 \,,
\ee
and the fixed points for the first, second and third twisted sectors are: 
\begin{eqnarray}
&& f^{(1)} = 0\,, \nonumber\\
&& f^{(2)}= 0, ~e_2/3, ~ 2e_2/3\,,   \nonumber\\
&& f^{(3)}= 0,~ e_2/2,  ~e_1/2, ~\frac{e_1+e_2}{2} \,.
\end{eqnarray}
Thus the physical states $|\psi^{(k)}\rangle$ for the $k$-th twisted sector are (see Eq. (\ref{psi})): 
\bea
&& |\psi^{(1)}\rangle = |0\rangle\,,  \nonumber\\
&& |\psi^{(2)} \rangle = |0\rangle,~|e_2/3\rangle  + e^{-2\pi\imath \gamma^{(2)}} |\theta (e_2/3) \rangle    \,,\nonumber  \\ 
&& |\psi^{(3)} \rangle = |0\rangle,~|e_1/2\rangle  + e^{-2\pi\imath \gamma^{(3)}} 
|\theta (e_1/2) \rangle +  e^{-4\pi\imath \, \gamma^{(3)}} |\theta^2 (e_1/2)\rangle   \,,
\eea
with $\gamma^{(2)} \in \{1/2, 1\}$ and $\gamma^{(3)} \in \{1/3, 2/3, 1\}$.

\bigskip

\FIGURE{
\includegraphics[width=13.25cm]{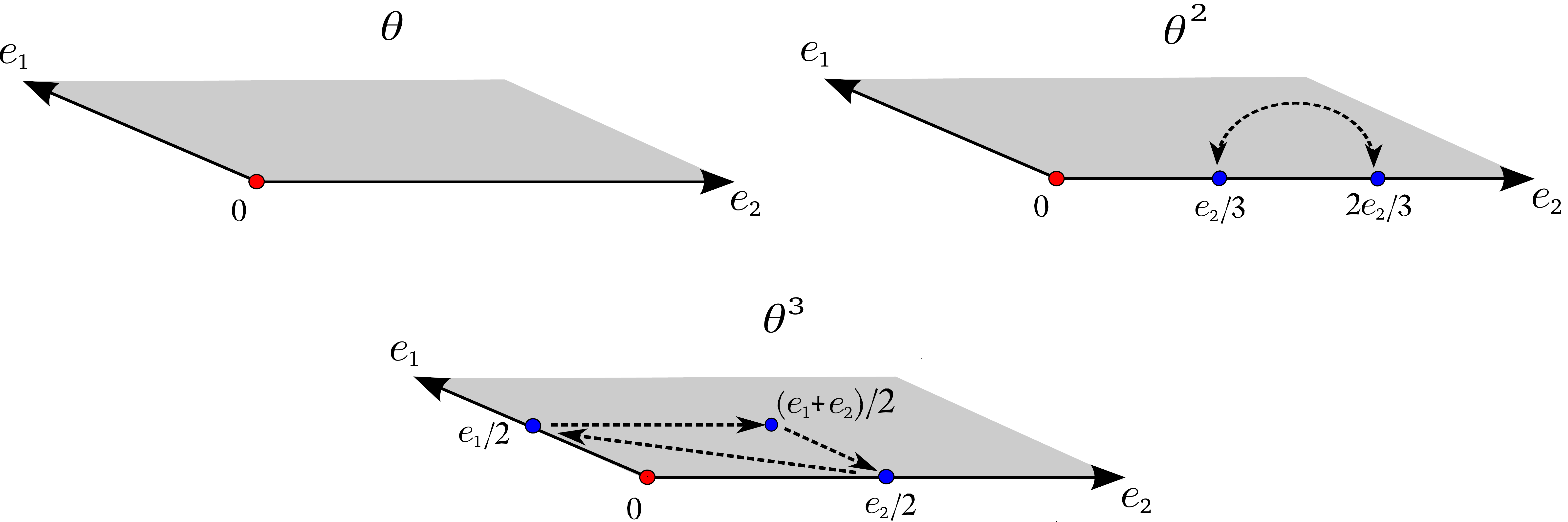}
\caption{Fixed point structure for the various twisted sectors of the
$T^2/\mathbbm{Z}_6$ orbifold. Note that the fixed points in blue are related by the orbifold twist, so they are equivalent in the orbifold space.}
\label{2DEjemplo}}

\bigskip

Consider for example a coupling between physical states of type $\theta^2\theta^2\theta^2$, say:
\be
|\psi_1^{(2)} \rangle|\psi_2^{(2)} \rangle|\psi_3^{(2)} \rangle=|0\rangle \left(|e_2/3\rangle  + e^{-2\pi\imath \gamma_2} |\theta (e_2/3) \rangle \right)\left(|e_2/3\rangle  + e^{-2\pi\imath \gamma_3} |\theta (e_2/3) \rangle\right) \,,
\ee
where we suppressed the twisted sector label to lighten the notation, so that $\gamma^{(2)}_\alpha= \gamma_\alpha$. 
The coupling is thus given by a sum of auxiliary couplings, weighted by the $\gamma$-phases, and differing only in the positions of the fixed points.  Applying the space group selection rule to each auxiliary coupling:
\be
(1-\theta^{k_3})f_3 + \theta^{k_3}(1-\theta^{k_2})f_2 + \theta^{k_3+k_2}(1-\theta^{k_1})f_1 \in \bigcup_{\alpha=1}^3 (1-\theta^{k_\alpha})\Lambda\,,
\ee
one finds that the auxiliary couplings $|0\rangle|e_2/3\rangle |e_2/3\rangle$ and $|0\rangle|\theta e_2/3\rangle |\theta e_2/3\rangle$ vanish.

Now we are ready to consider the correlation functions.  For a $\theta^2\theta^2\theta^2$ coupling only the holomorphic worldsheet instanton solutions are non-vanishing.  After applying the H- and gauge-momentum conservation and Rule 5, the correlation function can be non-vanishing provided $\calN_L \geq \bar\calN_L$, and the non-trivial part of each auxiliary correlation function takes the form (see Eq. (\ref{E:Faux})):
\be
{\mathcal F}_{aux} = \sum_{X_{cl}} e^{-S_{cl}} (\partial X_{cl})^{\calN_L-\bar\calN_L} \langle \sigma_{(\theta^2,f_1)} \sigma_{(\theta^2,f_2)} \sigma_{(\theta^2,f_3)} \rangle \,,
\ee
with the notation:
\be
\langle \sigma \dots \sigma \rangle = \int {\cal D}X_{qu} e^{-S_{qu}} \partial X_{qu}\dots \partial \bar X_{qu} \dots \sigma \dots \sigma \,,
\ee 
for the quantum part.  The quantum part constitutes a global factor for all the auxiliary correlation functions, as it does not depend on the positions of the fixed points, but only, via the local monodromy, to which twisted sectors they belong \cite{Dixon, twistfields, CGM}.  We are interested instead in the classical contribution, and in particular in the sum over worldsheet instanton solutions, which depends on the fixed points at which the twisted strings are localized.

The sum over classical solutions corresponds to a sum over lattice coset vectors, which for our couplings $|0\rangle|e_2/3\rangle |\theta e_2/3\rangle$ and $|0\rangle|\theta e_2/3\rangle |e_2/3\rangle$ are given, respectively, by (see Eq. (\ref{3ptcoset})):
\bea
\nu_1&=&(1-\theta^2)(e_2/3 -2e_1 - e_2 + \Lambda), \\
\nu_2&=&(1-\theta^2)(2e_2/3 - e_1-e_2 + \Lambda) \,.
\eea
Notice that these are related by a $\theta$-twist, $\theta\nu_1 = \nu_2$, as they must be since the fixed points are related by $\theta \left\{0,e_2/3,\theta e_2/3\right\} = \left\{0,\theta e_2/3,e_2/3\right\}$.  Moreover, it is easy to check that the sets of vectors $\nu_1$ and $\nu_2$ each have a $\Z_3$ rotation symmetry, generated by $\theta^2$, as they must since $\theta^2\left\{0,e_2/3,\theta e_2/3\right\}=\left\{0,e_2/3,\theta e_2/3\right\}$ and  $\theta^2\left\{0,\theta e_2/3,e_2/3\right\} = \left\{0,\theta e_2/3,e_2/3\right\}$.

We can use these relations to combine the auxiliary correlation functions:
\bea
\calF &=& e^{-2\pi\imath \gamma_3} \sum_{X_{cl}} e^{-S_{cl}} (\partial X_{cl})^{\calN_L-\bar\calN_L} \langle \sigma_{(\theta^2,0)} \sigma_{(\theta^2,e_1/3)} \sigma_{(\theta^2,\theta e_1/3)} \rangle \nonumber\\
&&\quad + e^{-2\pi\imath \gamma_2}  \sum_{X_{cl}} e^{-S_{cl}} (\partial X_{cl})^{\calN_L-\bar\calN_L} \langle \sigma_{(\theta^2,0)} \sigma_{(\theta^2,\theta e_1/3)} \sigma_{(\theta^2,e_1/3)} \rangle\,,
\eea
finding an overall factor:
\bea
\calF &\sim& e^{-2\pi\imath \gamma_3}\left((1)^{\calN_L-\bar\calN_L}+(\omega^2)^{\calN_L-\bar\calN_L} + (\omega^4)^{\calN_L-\bar\calN_L}\right)\nonumber\\
&&\quad + e^{-2\pi\imath \gamma_2}\left((\omega)^{\calN_L-\bar\calN_L} + (\omega^3)^{\calN_L-\bar\calN_L} + (\omega^5)^{\calN_L-\bar\calN_L} \right) \nonumber\\
&=& (\omega)^{-6\gamma_3}\left(1+\omega^{\calN_L-\bar\calN_L-6\sum_\alpha \gamma_\alpha} + \omega^{2\left(\calN_L-\bar\calN_L-6\sum_\alpha \gamma_\alpha\right)} + \omega^{3\left(\calN_L-\bar\calN_L-6\sum_\alpha \gamma_\alpha\right)}\right.\nonumber\\
&&\qquad\qquad\qquad \left. + \omega^{4\left(\calN_L-\bar\calN_L-6\sum_\alpha \gamma_\alpha\right)}+ \omega^{5\left(\calN_L-\bar\calN_L-6\sum_\alpha \gamma_\alpha \right)} \right) \,, \label{geomseries}
\eea
with $\omega=e^{2\pi\imath/6}$, where we recall that $\calN_L$ is shorthand for $\sum_\alpha \calN_{L\alpha}$, and $\gamma_{2,3}$ are shorthand for $\gamma_{2,3}^{(2)}$ so that $\gamma_{2,3}\in\{1/2,1\}$.
From the first line of Eq.~(\ref{geomseries}), by using the geometric series, it follows that correlation functions are vanishing unless:
\bea
\sum_{\alpha=1}^3 \calN_{L\,\alpha}-\bar\calN_{L\,\alpha} = 0 \mod 3 \,, \label{2drule6}
\eea
where we have reinstated the indices in the oscillator numbers. Similarly, from the second line of Eq.~(\ref{geomseries}), we require:
\bea
\sum_{\alpha=1}^3 \calN_{L\,\alpha}-\bar\calN_{L\,\alpha}-6\gamma_\alpha = 0 \mod 6 \,, \label{2dRcharge}
\eea
for non-vanishing correlation functions.
Notice that the above conditions (\ref{2drule6},\ref{2dRcharge}) are not independent in the simple case under consideration, but we will see below 6D examples where the analogous conditions are independent.

Couplings of the kind $\theta \theta^2 \theta^3$ can be analysed in a similar way.  Consider for example the coupling:
\be
|0\rangle \left(|e_2/3\rangle  + e^{-2\pi\imath \gamma_2} 
|\theta (e_2/3) \rangle \right) \left(   |e_1/2\rangle  + e^{-2\pi\imath \gamma_3} 
|\theta (e_1/2) \rangle +  e^{-4\pi\imath \, \gamma_3} |\theta^2 (e_1/2)\rangle \right) \,.
\ee
This time, the space group selection rule allows each of the auxiliary couplings.  The non-trivial part of the auxiliary correlation functions are:
\be
{\mathcal F}_{aux} = \sum_{X_{cl}} e^{-S_{cl}} (\partial X_{cl})^{\calN_L-\bar\calN_L} \langle \sigma_{(\theta,f_1)} \sigma_{(\theta^2,f_2)} \sigma_{(\theta^3,f_3)} \rangle \,,
\ee
with the sum over lattice coset vectors for $|0\rangle |e_1/3\rangle|e_1/2\rangle$, $|0\rangle |e_1/3\rangle|\theta e_1/2\rangle$, $|0\rangle |e_1/3\rangle|\theta^2 e_1/2\rangle$, $|0\rangle |\theta e_1/3\rangle|e_1/2\rangle$, $|0\rangle |\theta e_1/3\rangle|\theta e_1/2\rangle$, $|0\rangle |\theta e_1/3\rangle|\theta^2e_1/2\rangle$, given respectively by:
\bea
\nu_1 &=& (e_1/3 + e_1 + 2 \Lambda)\,, \nonumber\\
\nu_2 &=& (e_1/3 + 2 \Lambda)\,, \nonumber\\
\nu_3 &=& (e_1/3 -e_1 + e_2 +  2 \Lambda)\,, \nonumber\\
\nu_4 &=& (2e_1/3 + e_1 + e_2 + 2 \Lambda)\,, \nonumber\\
\nu_5 &=& (2e_1/3 + e_2 + 2 \Lambda)\,, \nonumber\\
\nu_6 &=& (2e_1/3 -e_1 + e_2 +  2 \Lambda) \,.
\eea
We observe that 
\be
\nu_1\rightarrow \nu_5 \rightarrow \nu_3 \rightarrow \nu_4 \rightarrow \nu_2\rightarrow \nu_6,
\ee
under the $\theta$ twist, as must be since the fixed points are related as:
\bea
&&\left\{0,e_1/3,e_1/2\right\} \rightarrow \left\{0,\theta e_1/3,\theta e_1/2\right\} \rightarrow \left\{0,e_1/3,\theta^2 e_1/2\right\} \nonumber\\
&&\qquad \rightarrow \left\{0,\theta e_1/3,e_1/2\right\}  \rightarrow\left\{0,e_1/3,\theta e_1/2\right\} \rightarrow
\left\{0,\theta e_1/3\,\theta^2e_1/2\right\} \,.
\eea
Then we find the correlation function can be written with an overall factor:
\bea
\calF &\sim& \left( (1)^{\calN_L-\bar\calN_L} + e^{-2\pi\imath \gamma_3}(\omega^4)^{\calN_L-\bar\calN_L} + e^{-4\pi\imath \gamma_3}(\omega^2)^{\calN_L-\bar\calN_L}+ e^{-2\pi\imath \gamma_2}(\omega^3)^{\calN_L-\bar\calN_L}\right.\nonumber\\
&&\qquad \left. + e^{-2\pi\imath \gamma_2 -2\pi\imath \gamma_3}(\omega)^{\calN_L-\bar\calN_L}  + e^{-2\pi\imath \gamma_2 -4\pi\imath \gamma_3}(\omega^5)^{\calN_L-\bar\calN_L} \right) \nonumber\\
&\sim&\left(1+\omega^{\calN_L-\bar\calN_L-6\sum_{\alpha} \gamma_{\alpha}} + \omega^{2\left(\calN_L-\bar\calN_L-6\sum_{\alpha} \gamma_{\alpha}\right)}+ \omega^{3\left(\calN_L-\bar\calN_L-6\sum_{\alpha} \gamma_{\alpha}\right)}\right.\nonumber\\
&&\qquad \left.+ \omega^{4\left(\calN_L-\bar\calN_L-6\sum_{\alpha} \gamma_{\alpha}\right)}+ \omega^{5\left(\calN_L-\bar\calN_L-6\sum_{\alpha} \gamma_{\alpha}\right)} \right) \,,
\eea
and hence the correlation functions are vanishing unless:
\be
\sum_{\alpha=1}^3\calN_{L\,\alpha}-\bar\calN_{L\,\alpha}-6 \gamma_\alpha = 0 \mod 6 \,.
\ee

Finally, we can make a similar analysis for the $\theta\theta\theta^4$ couplings, with the result:
\bea
&&\sum_{\alpha=1}^3 \calN_{L\,\alpha}-\bar\calN_{L\,\alpha} = 0 \mod 3\,, \\
&&\sum_{\alpha=1}^3 \calN_{L\,\alpha}-\bar\calN_{L\,\alpha}-6 \gamma_\alpha = 0 \mod 6 \,,
\eea
where again the two conditions are not independent.

In summary, we find that the symmetries relating the orbifold fixed points lead to symmetries among the worldsheet instanton solutions, which imply that all couplings are vanishing unless
\be
\sum_{\alpha=1}^3 \calN_{L\,\alpha}-\bar\calN_{L\,\alpha}-6 \gamma_\alpha = 0 \mod 6 \,. \label{G2Z6preRcharge}
\ee
If we combine this selection rule with the H-momentum conservation condition\footnote{A priori, for the 3-point couplings, we must still impose H-momentum conservation as an independent selection rule, see comment below Eq.~(\ref{Hmom}), and a further comment in what follows.} (\ref{Hmom}), we can write it as an R-charge conservation law:
\be
\sum_\alpha R_\alpha = 1 \mod 6 \quad \textrm{with} \quad R_\alpha = q_{sh\, \alpha} - \calN_{L\,\alpha}+\bar\calN_{L\alpha} + 6 \gamma_\alpha \,. \label{G2Z6Rcharge}
\ee
Importantly and not to be missed, the $\gamma$-phases contribute non-trivially to the R-charges.

Notice that in our simple 2D example, the orbifold geometrical symmetry under discussion is nothing more than the orbifold twist symmetry itself, and yet we are led to conditions that are in general independent of the orbifold projection condition, the 2D version of Eq. (\ref{GSO}), which reads:
\be
\sum_\alpha \calN_{L\,\alpha}-\bar\calN_{L\,\alpha}+6 \gamma_\alpha = 0 \mod 6 \,, \label{G2Z6GSO}
\ee
after application of gauge and H-momentum conservation and the space group selection rule\footnote{The sum of the vacuum phases is always trivial, as can be shown via modular invariance and the space group selection rule \cite{patrick}.}.
The orbifold projection condition, Eq. (\ref{G2Z6GSO}), is satisfied automatically for any combination of physical states.  In contrast, the R-charge conservation law, Eqs. (\ref{G2Z6preRcharge}) or (\ref{G2Z6Rcharge}), follows from the structure of the correlation function, and is due to the relations between fixed points or worldsheet instantons.  For prime orbifolds, where there is no need to build the physical states via the $\gamma$-phase, the  R-charge conservation law is identical to the orbifold twist invariance condition in 2D. 
They also happen to be identical for the $\Z_4$ case, since there we always have $\gamma \in \{0, 1/2\}$, so that $\gamma \simeq -\gamma$.  However, they are independent in general.

\subsection{The general 6D case}

With our simple 2D example in mind, let us now consider the general 6D case, where we can have additional symmetries beyond the point group, as enumerated in Subsection {\ref{S:lorentz}}.

A general L-point coupling between physical states,
\be
|\psi_\alpha\rangle = \sum_{r= 0}^{l_\alpha-1} 
\,e^{-2\pi\imath \,r\gamma_\alpha} |\theta^r f_\alpha \rangle \,,
\ee
can be written as a sum of auxiliary couplings, weighted by the $\gamma$-phases:
\be
\langle \psi_1 \dots \psi_\L \rangle = \sum_{r_1=0}^{l_1-1}\dots
\sum_{r_\L=0}^{l_\L-1} \,e^{-2\pi\imath(r_1\gamma_1 + \dots +
r_\L\gamma_\L)} \langle \theta^{r_1}f_1 \dots \theta^{r_\L}f_\L
\rangle \, ,
\ee
where we recall that $l_\alpha$ is the smallest integer such that $\theta^{l_\alpha} f_\alpha = f_\alpha + \lambda$, and $\gamma_\alpha = p_\alpha/l_\alpha$ for some integers $p_\alpha$ in the range $0, 1, \dots, l_\alpha-1$.

In order to avoid being distracted by clutter, let us assume to begin with that holomorphic instantons are allowed in all planes.  We will see by the end of the section that the general case follows straightforwardly.  As we discussed in Section \ref{S:calF} and Eq. (\ref{E:Faux}), the non-trivial part of the correlation function, which depends on oscillators and twist fields, takes the form:
\bea
{\mathcal F} &\sim& \sum_{r_1=0}^{l_1-1}\dots \sum_{r_\L=0}^{l_\L-1} e^{-2\pi\imath(r_1\gamma_1 + \dots + r_\L\gamma_\L)} \prod_{i=1}^3 \sum_{X_{cl}^i}\, e^{-S_{cl}^i} (\partial X_{cl}^i)^{({\mathcal N}_L^i-\bar{\mathcal N}_L^i-\bar{\mathcal N}_R^i)} \, \mathit{f}\left(|\partial X^i_{cl}|^2,\partial X_{cl}^i \partial \bar X_{cl}^i\right) \nonumber \\
&&\qquad\qquad\qquad\qquad\qquad\qquad\qquad\qquad \times  \langle \sigma_{(k_1,\theta^{r_1}f_1)}\dots\sigma_{(k_\L,\theta^{r_\L}f_\L)}\rangle\,,
\label{6dgenF} 
\eea
when holomorphic instantons are allowed, where again note that the classical solutions depend on the fixed points $\left\{\theta^{r_1} f_1, \dots, \theta^{r_\L} f_\L\right\}$ in each auxiliary correlation function.
We may now use this expression, and the relations among the fixed points and corresponding classical solutions, to derive selection rules.

\subsubsection{Factorizable orbifolds}

The orbifolds $T^6/\Z_3$, $T^6/\Z_4$, $T^6/\Z_{6-I}$ and $T^6/\Z_{6-II}$ may be constructed on factorizable lattices.
For these factorizable orbifolds, we can rewrite Eq. (\ref{6dgenF}) as:
\bea
{\mathcal F} &\sim& \sum_{r_1=0}^{l_1-1}\dots \sum_{r_\L=0}^{l_\L-1} e^{-2\pi\imath(r_1\gamma_1 + \dots + r_\L\gamma_\L)} \prod_{i=1}^3 \sum_{X_{cl}^i}\, e^{-S_{cl}^i} (\partial X_{cl}^i)^{({\mathcal N}_L^i-\bar{\mathcal N}_L^i-\bar{\mathcal N}_R^i)} \, \mathit{f}\left(|\partial X^i_{cl}|^2,\partial X_{cl}^i \partial \bar X_{cl}^i\right)\nonumber\\
&&\qquad\qquad\qquad\qquad\qquad\qquad\qquad\qquad \times \langle \sigma^i_{(k_1,\theta_i^{r_1}g_{1\,i})}\dots \sigma^i_{(k_\L,\theta_i^{r_\L}g_{\L\,i})}\rangle \,. \label{6dfactF}
\eea

Now, let us suppose that some planes, $j$, have prime ordered twists.  Then all the fixed points are $\theta$-invariant in $j$, that is, $\theta_j g_{\alpha\,j} = g_{\alpha\,j}$ for all $\alpha$.  It follows that the correlation function factorizes in prime planes:
\bea
{\mathcal F} &\sim& \prod_{j} \sum_{X_{cl}^j}\, e^{-S_{cl}^j} (\partial X_{cl}^j)^{({\mathcal N}_L^j-\bar{\mathcal N}_L^j-\bar{\mathcal N}_R^j)} \, \mathit{f}\left(|\partial X^j_{cl}|^2,\partial X_{cl}^j \partial \bar X_{cl}^j\right) \langle \sigma^j_{(k_1,g_{1\,j})} \dots \sigma^j_{(k_\L,g_{\L\,j})}  \rangle \nonumber \\
&&  \times \sum_{r_1=0}^{l_1-1}\dots \sum_{r_\L=0}^{l_\L-1} e^{-2\pi\imath(r_1\gamma_1 + \dots + r_\L\gamma_\L)} \prod_{i\neq j} \sum_{X_{cl}^i}\, e^{-S_{cl}^i} (\partial X_{cl}^i)^{({\mathcal N}_L^i-\bar{\mathcal N}_L^i-\bar{\mathcal N}_R^i)} \, \mathit{f}\left(|\partial X^i_{cl}|^2,\partial X_{cl}^i \partial \bar X_{cl}^i\right)\nonumber\\
&&\qquad\qquad\qquad\qquad\qquad\qquad\qquad\qquad\qquad\qquad \times \langle \sigma^i_{(k_1,\theta_i^{r_1}g_{1\,i})} \dots \sigma^i_{(k_\L,\theta_i^{r_\L}g_{\L\,i})}  \rangle \,. \label{6dprimefactF}
\eea
Moreover, from the observations at the end of Section \ref{S:calF}, it can be seen that the 
classical solutions $\left\{\partial X_{cl}^j\right\}$ to be summed over are related by $\theta_j$ twists:
\be
\theta_j: \quad \partial X_{cl\,1}^j \rightarrow \omega_j \partial X_{cl\,1}^j=\partial X_{cl\,2}^j\,, \quad \textrm{with } \omega_j=e^{2\pi\imath/\N^j}\,,
\ee 
where $\N^j$ is the order of the twist $\theta_j$.  The sum over classical solutions thus leads to an overall factor:
\be
\calF \sim \left((1)^{({\mathcal N}_L^j-\bar{\mathcal N}_L^j-\bar{\mathcal N}_R^j)} + (\omega_j)^{({\mathcal N}_L^j-\bar{\mathcal N}_L^j-\bar{\mathcal N}_R^j)} + \dots + (\omega_j^{(\N^j-1)})^{({\mathcal N}_L^j-\bar{\mathcal N}_L^j-\bar{\mathcal N}_R^j)} \right) \,,
\ee
which implies a selection rule for the prime planes:
\be
\sum_\alpha {\mathcal N}_{L\,\alpha}^{j} - \bar{\mathcal N}_{L\,\alpha}^{j} - \bar{\mathcal N}_{R\,\alpha}^{j} = 0 \mod \N^{j} \,. \label{E:twistinv}
\ee
Subsequently, the non-vanishing correlation functions are invariant under the orbifold twist  acting independently in the prime planes, which corresponds to a Lorentz symmetry in the orbifold geometry, as we saw in Subsection \ref{S:lorentz}.
This is the familiar  $\Z_{\N^{j}}$ twist invariance condition for prime 
planes, and combining with H-momentum conservation (\ref{Hmom}), it leads to the R-charge conservation law
\be
\sum_\alpha R^j_\alpha = 1 \mod \N^j\,, \quad \textrm{with  } R^j_\alpha = q_{sh\,\alpha}^j-\calN_{L\,\alpha}^j + \bar\calN_{L\,\alpha}^j \,. \label{E:Rfact}
\ee

To make further progress for the non-prime planes, consider a general coupling between some number $\L_1$ of twisted sectors associated with $l_1$ up to $\L_\L$ twisted sectors with $l_\L$, and take $\K$ the lowest common mutiple of $l_1, \dots, l_\L$, so $\theta^{\K}f_\alpha = f_\alpha + \lambda$ for all the associated fixed points $f_\alpha$.  The number of auxiliary couplings is $l_1^{\L_1} \times \dots \times l_\L^{\L_\L}$, and these can be divided into $(l_1^{\L_1} \times \dots \times l_\L^{\L_\L})/\K$ sets, each with $\K$ elements whose fixed points are related by powers of $\theta$ up to $\K-1$. 
For example, for a $\theta^2\theta^2\theta^2$ coupling in a $\Z_6$ orbifold, we have $\left\{\left\{ f_1, f_2, f_3 \right\}, \left\{ \theta f_1, \theta f_2, \theta f_3 \right\}\right\}$, $\left\{\left\{ f_1, f_2,  \theta f_3 \right\}, \left\{ \theta f_1, \theta f_2, f_3 \right\}\right\}$, $\left\{\left\{ f_1, \theta f_2, f_3 \right\}, \left\{ \theta f_1, f_2, \theta f_3 \right\}\right\}$ and $\left\{\left\{ \theta f_1, f_2, f_3 \right\}, \left\{ f_1 \theta f_2, \theta f_3 \right\}\right\}$.  If one element of a set is allowed by the space group selection rule, then all its elements are allowed. 
The classical solutions corresponding to the auxiliary correlation functions will be similarly related, so for example, for $\left\{ f_1, f_2, f_3 \right\}$ and $\left\{ \theta f_1, \theta f_2, \theta f_3 \right\}$ we have, respectively,
\be
\partial X_{cl}=(\partial X_{cl}^1,\,\partial X_{cl}^2, \,\partial X_{cl}^3) \quad \textrm{and} \quad \theta \partial X_{cl} = (e^{2\pi\imath v_1}\,\partial X_{cl}^1,\; e^{2\pi\imath v_2}\,\partial X_{cl}^2, \; e^{2\pi\imath v_3}\,\partial X_{cl}^3) \,.
\label{omegadX}
\ee 
Recall from Subsection \ref{S:lorentz} that we cannot factorize the $\theta$-twist plane by plane for non-prime planes.

Moreover, since all the fixed points are invariant under $\theta^\K$, it follows that the classical solutions corresponding to each auxiliary correlation function are related by powers of $\theta^{\K}$, up to\footnote{Notice that since all the $l_\alpha$ divide $\N$, by prime factorization so does their lowest common multiple $\K$.} $(\theta^{\K})^{\frac{\N}{\K}-1}$.  For the $\theta^2\theta^2\theta^2$ example, this corresponds to an additional $\Z_3$ symmetry, generated by $\theta^2$.  We can write this $\theta^{\K}$ symmetry among the classical solutions plane by plane (cf. Eq. (\ref{Rule6factorizable})), and thus the sum over classical solutions leads to an overall factor: 
\be
\calF_{aux} \sim \prod_{i\neq j}\left((1)^{({\mathcal N}_L^i-\bar{\mathcal N}_L^i-\bar{\mathcal N}_R^i)} + (\omega_i)^{({\mathcal N}_L^i-\bar{\mathcal N}_L^i-\bar{\mathcal N}_R^i)} + \dots + (\omega_i^{(\frac{\N^i}{\K}-1)})^{({\mathcal N}_L^i-\bar{\mathcal N}_L^i-\bar{\mathcal N}_R^i)} \right) \,, \label{omegaK/Ni}
\ee
with $\omega_i = e^{2\pi\imath\frac{\K}{\N^i}}$.  Therefore, in each of the planes with non-prime twist $\N^i$, we have the condition:
\be
\sum_\alpha {\mathcal N}_{L\,\alpha}^{i} - \bar{\mathcal N}_{L\,\alpha}^{i} - \bar{\mathcal N}_{R\,\alpha}^{i} = 0 \mod \frac{\N^i}{\K}. \label{rule6}
\ee

Next let us combine all the auxiliary correlation functions.  We need the relations amongst the classical solutions, but also amongst the $\gamma$-phase weightings. Consider for example an auxiliary coupling, with fixed points $\left\{ \theta^{r_1}f_1, \dots, \theta^{r_\L}f_\L \right\}$ that are related to the $\left\{f_1,f_2,\dots,f_\L\right\}$ by some twist $\theta^s$.  We have seen that the classical solutions mediating the first coupling are related to the second by a rotation $\theta^s$, and that there are $\K$ such couplings, with $s=0,\dots,\K-1$.  Now, the integer $s$ is related to the integers $r_\alpha$ as $\theta^{r_\alpha} f_\alpha = \theta^s f_\alpha + \lambda$ for all $\alpha$, that is $s-r_\alpha = m l_\alpha$ with integer $m$. Thus, the $\gamma$-phase weighting factor can be written as
\be
e^{-2\pi\imath(r_1\gamma_1 + \dots + r_\L\gamma_\L)}  = e^{-2\pi\imath s\sum_\alpha \gamma_\alpha}  \,,
\ee
where we have used that $\gamma_\alpha$ is an integer multiple of $1/l_\alpha$. 
Moreover, as we discussed above, if the fixed points $\left\{f_1,f_2,\dots, f_\L\right\}$ are all invariant under $\theta^{\K}$, then the classical solutions in each auxiliary coupling have a $\theta^{\K}$ symmetry.  These do not come with $\gamma$-phases weighting their relative contributions, but we can use $\K=m l_\alpha$ with integer $m$ for all $\alpha$, to insert factors of $1=e^{-2\pi\imath\K\sum_\alpha \gamma_\alpha}$.

Let us now apply all these relations to simplify the non-prime part of the full correlation function (\ref{6dprimefactF}).  We may put together the classical solutions from each set of auxiliary correlation functions that are related by $1,\theta, \dots, \theta^{\K-1}$, and moreover from within each auxiliary correlation function itself, which are related by $1, \theta^\K, \dots, (\theta^\K)^{\frac{\N}{\K}-1}$.  Combined, the classical solutions are all related by $1, \theta, \dots, \theta^{\N-1}$.  Taking care of the $\gamma$-phase weightings, one thus finds that the correlation function can be written with an overall factor:

\bea
{\mathcal F}\sim \prod_{i\neq j} \sum_{|X^i_{cl}|}\sum_{n=0}^{N-1}  e^{-S^i_{cl}}(|\partial X_{cl}^i|\omega^n)^{\N v^i({\mathcal N}_L^i-\bar{\mathcal N}_L^i-\bar{\mathcal N}_R^i)} \,e^{-2\pi\imath\,n\sum_{\alpha=1}^\L \gamma_\alpha}  \,,
\eea 
where $\omega=e^{2\pi\imath/\N}$, and we have split the sum over classical solutions into a sum over phases and lengths. From here one concludes immediately that correlation functions vanish unless:
\be
\sum_\alpha \left(\sum_{i\neq j} \N v^i \left({\mathcal N}_{L\,\alpha}^{i} - \bar{\mathcal N}_{L\,\alpha}^{i} - \bar{\mathcal N}_{R\,\alpha}^{i}\right) - \N \gamma_\alpha \right) = 0 \mod \N \,. \label{factnonprimeR}
\ee

This rule holds for all couplings, and putting together with H-momentum conservation (\ref{Hmom}), it can be expressed as an R-charge conservation law:
\be
\sum_\alpha R_\alpha = \left(\sum_{i\neq j}\N v^i\right) \mod \N\, \quad \textrm{with } \;R_\alpha = \sum_{i\neq j}\N v^i  \left(q_{sh \alpha}^i - {\mathcal N}_{L\,\alpha}^{i} + \bar{\mathcal N}_{L\,\alpha}^{i} \right) + \N \gamma_\alpha \,, \label{E:Rnonprime}
\ee
Observe that the correct R-charge for non-prime planes is summed over the non-prime planes, and has a non-trivial contribution from the $\gamma$-phase.

Notice that with at most one non-prime plane, the condition (\ref{rule6}) follows automatically after imposing the R-charge conservation law or (\ref{factnonprimeR}).  However, when there are two non-prime planes, the condition (\ref{rule6}) becomes an independent selection rule, which we may call Rule 6 or the {\it coset vector selection rule}.  
Since this condition depends on the relative properties of the twisted sectors involved in the couplings, namely the lowest common multiple of the twisted sectors (or their conjugates),  it cannot be interpreted as a conventional symmetry like R-symmetry in the 4D theory.  It is however a stringy selection rule, coming from the symmetries in the fixed points or worldsheet instantons.

To complete our analysis, so far in deriving the rules we assumed for simplicity that holomorphic instantons were allowed in every plane.  In general, holomorphic instantons may be forbidden in some (or all) planes, in which case the correlation functions also include factors of the form Eq. (\ref{E:Faux2}). The classical solutions to be summed over in the correlation functions are related by the $\theta$ twist as follows.  For $\left\{ f_1, f_2, f_3 \right\}$ and $\left\{ \theta f_1, \theta f_2, \theta f_3 \right\}$ we have, respectively,
\bea
&&\partial X_{cl}=(\partial X_{cl}^1,\,\partial X_{cl}^2, \,\partial X_{cl}^3)\,,\nonumber \\ 
&&\partial \bar X_{cl}=(\partial \bar X_{cl}^1,\,\partial \bar X_{cl}^2, \,\partial \bar X_{cl}^3)\,,
\eea
and
\bea
&&\theta \partial X_{cl} = (e^{2\pi\imath v_1}\,\partial X_{cl}^1,\; e^{2\pi\imath v_2}\,\partial X_{cl}^2, \; e^{2\pi\imath v_3}\,\partial X_{cl}^3) \,,\nonumber \\ 
&&\theta \partial \bar  X_{cl} = (e^{-2\pi\imath v_1}\,\partial \bar  X_{cl}^1,\; e^{-2\pi\imath v_2}\,\partial \bar  X_{cl}^2, \; e^{-2\pi\imath v_3}\,\partial \bar  X_{cl}^3) \,.
\eea
From here, it follows that the selection rules in the general case take the same form found above, Eqs. (\ref{E:Rfact}), (\ref{rule6}), (\ref{E:Rnonprime}).

Finally, we have already discussed how H-momentum conservation is not a selection rule for higher order couplings.  Following an analogous proof to that found in \cite{KPRZ}, it can be shown that even for three-point couplings, it is sufficient to impose R-charge conservation and the point group selection rule, to ensure that H-momentum conservation is automatically satisfied.

We are now ready to put all we have learned together, to write down the instanton rules for some representative examples.  We refer to \cite{Kobayashi:1991rp,CGM} for details on the torus lattices, twist actions and fixed points.

\paragraph{$T^6/\Z_{3}$ on $SU(3)\times SU(3) \times SU(3)$ lattice.}
Since this is a prime orbifold, there is no need to build the physical states via the $\gamma$-phases.  The correlation function factorizes into all three complex planes, and the R-charge conservation laws are as expected 
\cite{Rule4}:
\bea
&&\sum_\alpha R^i_\alpha = 1 \mod 3\, \quad \textrm{with } R^i_\alpha =  q_{sh\,\alpha}^i - \calN_{L \,\alpha}^i + \bar\calN_{L \,\alpha}^i\,, 
\qquad i = 1, 2, 3  \,.
\eea
Notice that, for the prime factorizable orbifold, we have an R-charge condition for each plane.
The sum of the three conditions is equivalent to the orbifold projection condition, which is automatically satisfied for all possible combinations of states.

\paragraph{$T^6/\Z_{4}$ on $SO(4)\times SO(4) \times SO(4)$ lattice.}
Only the third plane is prime, allowing us to factorize the correlation function there.  The R-charge conservation law is:
\bea
&&\sum_\alpha R_\alpha = 2 \mod 4\,, \quad \textrm{with } R_\alpha = \sum_{i=1}^2 \left(q_{sh\,\alpha}^i - \calN_{L \,\alpha}^i + \bar\calN_{L \,\alpha}^i \right)+ 4\gamma_\alpha\,,\\
&&\sum_\alpha R^3_\alpha = 1 \mod 2\,, \quad \textrm{with } R^3_\alpha = q_{sh\,\alpha}^3 - \calN_{L \,\alpha}^3 + \bar\calN_{L \,\alpha}^3 \quad \textrm{(not independent)}  \,,
\eea
and moreover
\bea
&&\sum_\alpha R^1_\alpha = 1 \mod 2\,, \quad \textrm{with } R^1_\alpha = q_{sh\,\alpha}^1 - \calN_{L \,\alpha}^1 + \bar\calN_{L \,\alpha}^1\,,\\
&&\sum_\alpha R^2_\alpha = 1 \mod 2\,, \quad \textrm{with } R^2_\alpha = q_{sh\,\alpha}^2 - \calN_{L \,\alpha}^2 + \bar\calN_{L \,\alpha}^2   \quad \textrm{(not independent)}\,,
\eea
since all states are  twisted under either $\theta$ or $\theta^2$.

\paragraph{$T^6/\Z_{6-I}$ on $G_2\times G_2 \times SO(4)$ lattice.}
Again, only the third plane is prime, and so the correlation function in general only factorizes there.  The R-charge conservation laws are:
\bea
&&\sum_\alpha R^3_\alpha = 1 \mod 3\,, \quad \textrm{with } R_\alpha^3 = q^3_{sh\,\alpha} - \calN_{L \,\alpha}^3 + \bar\calN_{L \,\alpha}^3\,, \\
&&\sum_\alpha R_\alpha = 2 \mod 6\,, \quad \textrm{with } R_\alpha =  \sum_{i=1}^2\left( q^i_{sh\,\alpha} - \calN_{L \,\alpha}^i +\bar\calN_{L \,\alpha}^i \right)+6\gamma_\alpha  \,.
\eea

Moreover, for couplings involving only $\theta,\theta^2$ (and untwisted) sectors, we have the additional stringy Rule 6:
\bea
&&\sum_\alpha \calN_{L \,\alpha}^1 - \bar\calN_{L \,\alpha}^1-\bar\calN_{R \,\alpha}^1  = 0 \mod 3\,,\\
&&\sum_\alpha \calN_{L \,\alpha}^2 - \bar\calN_{L \,\alpha}^2-\bar\calN_{R \,\alpha}^2  = 0 \mod 3 \qquad \textrm{(not independent)}\,,
\eea
and for couplings involving only $\theta,\theta^3$ (and untwisted) sectors, we have:
\bea
&&\sum_\alpha \calN_{L \,\alpha}^1 - \bar\calN_{L \,\alpha}^1-\bar\calN_{R \,\alpha}^1  = 0 \mod 2\,,\\
&&\sum_\alpha \calN_{L \,\alpha}^2 - \bar\calN_{L \,\alpha}^2-\bar\calN_{R \,\alpha}^2  = 0 \mod 2 \qquad \textrm{(not independent)}\,.
\eea
As these rules are coupling dependent, they cannot be interpreted as a charge conservation law and symmetry in the low energy effective field theory.

\paragraph{$T^6/\Z_{6-II}$ on $G_2\times SU(3) \times SO(4)$ lattice.}
This orbifold has only one non-prime plane, and hence the correlation function factorizes fully as in the $\Z_3$ case.  There are three independent 
R-charge conservation laws:
\bea
&&\sum_\alpha R^1_\alpha = 1 \mod 6\, \quad \textrm{with } R^1_\alpha = q_{sh\,\alpha}^1 - \calN_{L \,\alpha}^1 + \bar\calN_{L \,\alpha}^1+ 6\gamma_\alpha\,, \\
&&\sum_\alpha R^2_\alpha = 1 \mod 3\, \quad \textrm{with }  R^2_\alpha = q_{sh\,\alpha}^2 - \calN_{L \,\alpha}^2 + \bar\calN_{L \,\alpha}^2\,, \\
&&\sum_\alpha R^3_\alpha = 1 \mod 2\, \quad \textrm{with }  R^3_\alpha = q_{sh\,\alpha}^3 - \calN_{L \,\alpha}^3 + \bar\calN_{L \,\alpha}^3
\,.
\eea
The $\gamma$-phase contributes non-trivially to the R-charges of the non-prime plane.    
The conditions are independent of the orbifold projection condition.

\subsubsection{Non-factorizable orbifolds}
\label{S:nonfact}
As we observed in Subsection \ref{S:lorentz}, for the non-factorizable orbifolds the twist does not act plane by plane.  However, by definition, the fixed points are invariant under, or related by, the twist acting in all the planes at the same time.  Moreover, in some cases, we found additional symmetries that leave the fixed points invariant.  

Let us first consider the consequences of the orbifold twist symmetry in both the prime and non-prime cases.  For the prime cases, this leads only to the orbifold projection condition, which is satisfied by all possible combinations of physical states.  Thus we have no R-charge conservation law or Rule 6 for the non-factorizable, prime orbifolds. However, for the non-prime orbifolds, the orbifold twist does give rise to  additional selection rules.  The derivation is exactly analogous to the factorizable case, but now, as the twist does not act plane by plane, all the planes must be taken together.  

In detail, consider a coupling between twisted sectors that are all fixed under $\theta^\K$.  The set of classical instanton solutions $\partial {X}_{cl}$ for each auxiliary function enjoys the discrete symmetry generated by $\theta^\K$, with
\be
\theta^\K \partial X_{cl} = (e^{2\pi\imath \K v^1}\, \partial X^1_{cl},\; e^{2\pi\imath \K v^2} \,\partial X^2_{cl},\;e^{2\pi\imath \K v^3} \partial X^3_{cl})\,,
\ee
but now it cannot be written plane by plane as in Eq. (\ref{omegaK/Ni}). 
Instead, each auxiliary correlation function has an overall factor:
\bea
\calF_{aux} &\sim& \left((1)^{\sum_{i=1}^3 \N v^i(\calN_L^i-\bar\calN_L^i-\bar\calN_R^i)} + (\omega)^{\sum_{i=1}^3 \N v^i (\calN_L^i-\bar\calN_L^i-\bar\calN_R^i)} + \dots \right. \nonumber \\
&& \qquad \qquad \dots \left. + (\omega^{(\frac{\N}{\K}-1)})^{\sum_{i=1}^3 \N v^i (\calN_L^i-\bar\calN_L^i-\bar\calN_R^i)}  \right)\,,
\eea
with $\omega=e^{2\pi\imath \frac{\K}{\N}}$.  From this expression 
the condition:
\be
{\sum_{i=1}^3 \N v^i ({\mathcal N}_L^i-\bar{\mathcal N}_L^i-\bar{\mathcal N}_R^i)} = 0 \mod \frac{\N}{\K}\,, \label{nonfactrule6}
\ee
immediately follows.

Now, in a similar way, by putting together the classical solutions from all the auxiliary correlation functions, we can derive a {\em $Q$-charge} conservation condition:
\be
\sum_\alpha Q_\alpha = 0 \mod \N\,, \quad \textrm{with } \quad Q_\alpha = \sum_{i=1}^3 \N v^i \left(q_{sh \,\alpha}^i   - {\mathcal N}_{L\,\alpha}^{i} + \bar{\mathcal N}_{L\,\alpha}^{i} \right) + \N \gamma_\alpha \,. \label{nonfactnonprimeR}
\ee
Using the twist vectors given in Table \ref{T:ZN}, and the relation between H-momentum in the spinor and vector representations (\ref{qsh}), one sees that the charges $Q_\alpha$ are the same for bosonic and fermionic superpartners. Hence, the charges $Q_\alpha$ do not correspond to R-charges. However, their conservation does constitute a non-trivial coupling selection rule, independent of the orbifold projection condition, and corresponding to a symmetry in the low energy effective field theory.

Finally, one can use the relation $\gamma_\alpha \K =\textrm{integer}$ to show that the condition (\ref{nonfactrule6}) is always satisfied once the charge conservation law (\ref{nonfactnonprimeR}) is imposed, so Rule 6 is trivial for the non-factorizable orbifolds.

Let us then give some examples, including some orbifolds whose underlying lattice and twist action is partially factorizable.

\paragraph{$T^6/\Z_{7}$ on $SU(7)$ lattice.}
This orbifold is both non-factorizable and prime, and hence we expect no Q- or R-charge selection rule and no Rule 6.

\paragraph{$T^6/\Z_{6-II}$ on $SU(6) \times SU(2)$ lattice.}
This orbifold is non-factorizable, but it is non-prime.  Hence, we derive an Q-charge conservation rule:
\be
\sum_\alpha Q_\alpha = 0 \mod 6\,, \quad \textrm{with } Q_\alpha  =  \sum_{i=1}^3 6 v^i \left(q_{sh\,\alpha}^i - \calN_{L \,\alpha}^i + \bar\calN_{L \,\alpha}^i \right) +6\gamma_\alpha  \,,
\ee
which is different to the orbifold projection condition.

\paragraph{$T^6/\Z_{8-II}$ on $SO(8) \times SO(4)$ lattice.}
This orbifold is partially factorizable, as both the underlying lattice and orbifold twist can be factorized on the third complex plane, where the twist is prime order 2.  Thus we obtain the R-charge conservation rules:
\bea
&&\sum_\alpha R_\alpha = 4 \mod 8\,, \quad \textrm{with } R_\alpha =  \sum_{i=1}^2 8 v^i \left(q_{sh\alpha}^i -\calN_{L \,\alpha}^i + \bar\calN_{L \,\alpha}^i\right)  +8\gamma_\alpha\,,\\
&&\sum_\alpha R_\alpha^3 = 1 \mod 2\,, \quad \textrm{with } R_\alpha^3 = q_{sh\,\alpha}^3 -   \calN_{L \,\alpha}^3 + \bar\calN_{L \,\alpha}^3 \,.
\eea

We can write down the rules for the $T^6/\Z_{12-I}$ orbifold on $SU(3)\times F_4$ and $T^6/\Z_{12-II}$ on $F_4\times SO(4)$ analogously.

\bigskip
Now let us consider the additional $\Z_2$ Lorentz symmetry, generated by $(\theta_1)^2$, which we observed in Subsection \ref{S:lorentz} for the $\Z_4$ orbifold on $SU(4)\times SU(4)$ and the $\Z_{8-I}$ orbifold on $SO(5) \times SO(9)$.  In detail, this $\Z_2$ acts in the first complex plane, and leaves all the fixed points invariant up to lattice shifts.  Consequently, the worldsheet instanton solutions for each auxiliary correlation function will enjoy the same symmetry, and this leads to an R-charge conservation law.  Then, we can write the selection rules for these orbifolds as follows.

\paragraph{$T^6/\Z_4$ on $SU(4)\times SU(4)$ lattice.}
The torus lattice and orbifold twist cannot be factorized onto the three complex planes in this example.  Therefore, a Q-charge conservation law  emerges from the non-prime orbifold twist:
\be
\sum_\alpha Q_\alpha = 0 \mod 4\,, \quad \textrm{with } Q_\alpha = \sum_{i=1}^3 4 v^i \left(q_{sh\,\alpha}^i - \calN_{L \,\alpha}^i + \bar\calN_{L \,\alpha}^i\right) +4\gamma_\alpha \,.
\ee
Meanwhile, the additional $\Z_2$ symmetry, generated by $(\theta_1)^2$, enjoyed by all the fixed points leads to an R-charge conservation law:
\be
\sum_\alpha R^1_\alpha = 1 \mod 2\,, \quad \textrm{with } R^1_\alpha =  \left(q_{sh\,\alpha}^1 - \calN_{L \,\alpha}^1 + \bar\calN_{L \,\alpha}^1\right) \,.
\ee

\paragraph{$T^6/\Z_{8-I}$ on $SO(5)\times SO(9)$ lattice.}
Here, although the torus lattice and orbifold twist can be factorized in the first complex plane, as the twist in that plane is non-prime, the correlation function does not factorize.  Therefore, a Q-charge conservation law emerges from the orbifold twist:
\be
\sum_\alpha Q_\alpha = 0 \mod 8\,, \quad \textrm{with } Q_\alpha = \sum_{i=1}^3 8 v^i \left(q_{sh\,\alpha}^i - \calN_{L \,\alpha}^i + \bar\calN_{L \,\alpha}^i\right) +8\gamma_\alpha\,.
\ee
Again, the additional $\Z_2$ symmetry generated by $(\theta_1)^2$ leads to an R-charge conservation law:
\be
\sum_\alpha R^1_\alpha = 1 \mod 2\,, \quad \textrm{with } R^1_\alpha = q_{sh\, \alpha}^1 -  \calN_{L \,\alpha}^1 + \bar\calN_{L \,\alpha}^1 \,.
\ee

\section{Rule 4 in Factorizable and Non-Factorizable Orbifolds}

Finally, we should reconsider Rule 4 for the factorizable and non-factorizable, prime and non-prime orbifolds.  The usual statement of Rule 4 for factorizable orbifolds is as follows \cite{Rule4}.  When all the twisted states in a coupling lie at the same fixed point in a given plane, the 
symmetries relating the worldsheet instanton solutions are the full torus 
lattice automorphisms, which may be larger than the orbifold twist $\N^i$.  This leads to an additional constraint for the non-vanishing couplings:
\be
\sum_\alpha \calN_{L \,\alpha}^i - \bar\calN_{L \,\alpha}^i-\bar\calN_{R \,\alpha}^i = 0 \mod \M^i\,,
\ee
where $\M^i$ is the order of the torus lattice $\Z_{\M^i}$ automorphism group in the $i$-th plane.  In the light of our previous discussion, this version of Rule 4 holds for factorizable prime planes.  For non-prime planes in factorizable orbifolds, it must be checked that the twisted fields are at the same fixed point in all the auxiliary couplings allowed by the space group selection rule.  In that case, every auxiliary coupling has the $\Z_{\M^i}$ symmetry in the $i$-th plane, so Rule 4 again applies as above, and is generally stronger than the R-charge conservation condition (\ref{factnonprimeR}), even when the order of the torus lattice automorphisms matches that of the point group.

For the non-factorizable orbifolds, when all twisted fields are at the same fixed point in each allowed auxiliary coupling, 6D torus (sub)lattice automorphisms such as the orbifold twist\footnote{In contrast, for couplings involving twisted fields at different fixed points in non-prime orbifolds, the orbifold twist symmetry is observed in the correlation function only after putting all the auxiliary correlation functions together.  The consequences of this symmetry were discussed in Section \ref{S:nonfact}.} $\theta$, are observed in each auxiliary correlation function.  Thus, we are 
lead to the condition (for all fields at the same fixed point in all auxiliary correlation functions): 
\be
\sum_{i=1}^3\sum_\alpha \N v^i (\calN_{L \,\alpha}^i - \bar\calN_{L \,\alpha}^i-\bar\calN_{R \,\alpha}^i) = 0 \mod \N \,,
\ee
which is independent of the orbifold projection condition for
non-prime orbifolds, and hence constitutes a non-trivial selection
rule there. 

Moreover, every torus lattice has a $\Z_2$ symmetry, that is, the lattice vectors always come in pairs $\left\{\lambda,-\lambda\right\}$.  
Thus we obtain the additional condition (again, for all fields at the same fixed point in all auxiliary correlation functions)\footnote{Note that the torus lattice has the $\Z_\N$ symmetry and the $\Z_2$ symmetry, but this does not necessarily imply it has a $\Z_{2\N}$ symmetry nor that $\Z_2 \subset \Z_\N$ for $\N$ even!}:
\be
\sum_{i=1}^3\sum_\alpha  (\calN_{L \,\alpha}^i - \bar\calN_{L \,\alpha}^i-\bar\calN_{R \,\alpha}^i) = 0 \mod 2 \,.
\ee

For example, for the $\Z_7$ orbifold on an $SU(7)$ root lattice, we have 
\be
\sum_{i=1}^3\sum_\alpha(\calN_{L \,\alpha}^i - \bar\calN_{L \,\alpha}^i-\bar\calN_{R \,\alpha}^i) = 0 \mod 2, 
\ee
whereas for a $\Z_{6-II}$ orbifold on an $SU(6)\times SU(2)$ lattice we have:
\bea
&&\sum_{i=1}^3\sum_\alpha 6 v^i(\calN_{L \,\alpha}^i - \bar\calN_{L \,\alpha}^i-\bar\calN_{R \,\alpha}^i) = 0 \mod 6\,, \\
&&\sum_{i=1}^3\sum_\alpha(\calN_{L \,\alpha}^i - \bar\calN_{L \,\alpha}^i-\bar\calN_{R \,\alpha}^i) = 0 \mod 2 \,.
\eea

Finally, we should note that as Rule 4 applies only to some couplings, in fact depending on the relative distance between the twisted strings in the orbifold space, it cannot be interpreted as a conventional symmetry in the low energy effective field theory \cite{Rcharge&miracle}.

\section{Conclusions}

In this paper we have derived string coupling selection rules for $T^6/\Z_\N$ orbifolds.  Our method builds on that used in \cite{KPRZ}, but we now consider the various classes of orbifolds,  with the torus lattice $\Lambda$ factorizable and non-factorizable, and the orbifold twist $\theta$ of prime and non-prime order.  We have found corrections to the old R-charge conservation rule for factorizable orbifolds, studied selection rules for non-factorizable orbifolds, and identified a new stringy selection rule that we call Rule 6.

Couplings between twisted strings generically have classical contributions from worldsheet instantons stretching between the fixed points and wrapping the orbifold geometry.  In computing the coupling strength, or corresponding correlation function, we have to sum over all such possible classical solutions.  The set of possible instanton solutions, which are proportional to vectors lying on particular cosets of the torus lattice, enjoys some symmetry.  This symmetry often leads some couplings to vanish.  In particular, we can write down some {\it instanton selection rules}, which turn out to constrain the number of oscillators that can appear in non-vanishing couplings.

One of these instanton selection rules is the well-known R-charge conservation law.  Actually, R-charge conservation is usually understood in terms of discrete Lorentz symmetries that survive the orbifolding.  In particular, in \cite{Rule4,Rcharge&miracle}, it was understood that the factorizable $T^6/\Z_3$ orbifold's geometry is invariant under the orbifold twist acting independently in each of the planes.  It turns out that the allowed couplings are similarly invariant under such independent twists, which  correspond to symmetries among the set of worldsheet instantons. Thus one is lead to a twist invariance condition, which, put together with H-momentum conservation, leads to a conserved R-charge.  In this way, discrete Lorentz symmetries in the orbifold geometry, which distinguish between bosons and fermions, lead naturally to discrete R-symmetries in the low energy effective field theory describing the compactification.

This intuition was subsequently applied to all factorizable orbifolds.  Hence the current consensus is that in factorizable orbifolds, there is a twist symmetry acting independently plane by plane, which leads to an R-charge conservation condition in each of the three planes, $\sum_\alpha R^i_\alpha = 1 \mod \N^i$, with R-charges given by $R^i_\alpha = q_{sh \, \alpha}^i - \calN^i_{L\,\alpha} + \bar\calN_{L\,\alpha}^i$ and $\N^i$ the order of the orbifold twist in the $i$-th plane.  Then it remained to understand possible R-charge conservation laws in non-factorizable orbifolds.  By carefully considering the derivation of the R-charge conservation rule from the orbifold CFT, however, we find the old R-charge conservation law to hold only for the prime, factorizable orbifold.

More generally, one has to take care when assigning the conserved R-charges to the string states.  
The string orbifold compactification includes twisted sectors, which are localized at the fixed point singularities and emerge only due to the compactification.  As a consequence, intuition gained from field theoretic smooth compactifications has to be checked in string orbifold compactifications.  Indeed, it is not clear how the twisted states transform under the orbifold twist acting independently in each plane.  Therefore, to identify the R-charge conservation law for allowed couplings, we have to consider the consequences of the orbifold's geometrical symmetries in the correlation functions, and in particular in the worldsheet instanton solutions.

For non-prime orbifolds, the higher twisted sectors are attached to fixed points that are not necessarily $\theta$-invariant.  Physical states are then built up using linear combinations of twisted strings attached to different fixed points in the same conjugacy class, whose constant coefficients  are powers of $e^{-2\pi\imath \gamma}$, with $\gamma$ the $\gamma$-phase \cite{physical}.   The geometrical part of the state is then a $\theta$-eigenstate, with eigenvalue $e^{2\pi\imath \gamma}$, and only the complete state is $\theta$-invariant. Correlation functions between physical states are then sums of auxiliary correlation functions between states at particular fixed points, weighted by the $\gamma$-phases.  Consequently, the sum over instanton solutions is also weighted by the $\gamma$-phases.  The twist relations among the fixed points do lead to symmetries among the instanton solutions, and these symmetries in the geometry do 
lead to R-charge conservation laws,  
but the conserved R-charges are different to those found in the literature.  In particular, the
$\gamma$-phase contributes non-trivially to the charges, and an independent R-charge conservation law is obtained only for prime planes in factorizable orbifolds, whereas non-prime planes all contribute to a single R-charge conservation. 

At this point, it may be useful to recall the rather convoluted history of the $\gamma$-phase in orbifold selection rules.  For several years it was believed that non-vanishing couplings must satisfy the so-called $\gamma$-rule, which stated that the sum of the $\gamma$-phases must be trivial \cite{CGM, Rcharge}.  Then, in \cite{nogammarule} it was emphasised that the $\gamma$-rule is in fact automatically satisfied by all couplings between physical states, after imposing gauge invariance and the R-charge conservation rule.  Careful  derivation of the selection rules from the CFT tells a different story.  R-charge conservation and H-momentum conservation constrain the sum of the $\gamma$-phases to be trivial for three-point couplings between massless ground states, that is, couplings that do not involve oscillators. However, there is no independent $\gamma$-rule, and in general R-charge conservation poses weaker constraints on the $\gamma$-phase.

A similar analysis can be made for non-factorizable orbifolds.  As the twist does not act plane by plane, there are in general no independent twist symmetries.  For prime orbifolds, the 6D twist symmetry that relates the worldsheet instanton solutions is automatically satisfied for all possible couplings, and thus there is no corresponding selection rule.  For non-prime orbifolds, the weighting of the worldsheet instanton solutions by the $\gamma$-phases in the correlation function leads to a single charge conservation law, where again, the $\gamma$-phases contribute to the charges, and now the corresponding symmetry is one that commutes with supersymmetry.  Moreover, in some special cases, the worldsheet instantons for non-factorizable orbifolds enjoy a further $\Z_2$ symmetry in one complex plane, which leads to an R-charge conservation law. 

Additional symmetries in the worldsheet instanton solutions can lead to more selection rules, which constrain the oscillator numbers.  An example of this is already known; Rule 4 or the {\it torus lattice selection rule}  \cite{Rule4,KPRZ}.  This occurs for couplings with all twisted fields at the same fixed point, in which case the symmetry among the worldsheet instanton solutions  is enhanced from the twist symmetry to the full torus lattice symmetries.  This applies, in different forms, for factorizable and non-factorizable, prime and non-prime orbifolds.  We have also found another such rule for the factorizable 
non-prime orbifolds, with two non-prime planes, which we may call Rule 6 or the {\it coset vector selection rule}.  When the lowest common multiple, $\K$, of the twisted sectors in a given coupling is less than the order $\N$ of the point group, the worldsheet instanton solutions enjoy an additional symmetry, of order $\N/\K$.   We summarize our results with Table \ref{T:rules}.  In Table \ref{T:Z6II}, we compare the number of allowed trilinear couplings in a $T^6/Z_{6-II}$ model\footnote{We computed the spectrum of this model using the orbifolder \cite{orbifolder}. Then we developed a code to compute the couplings implementing Rule 4, Rule 5, Rule 6 and the correct R-charge selection rules.}, applying the presently used selection rules with old R-charges, the presently used selection rules with correct R-charges, and the presently used selection rules with correct R-charges plus Rules 4, 5 and 6.
Observe that many couplings that were ruled out by the old R-charge conservation law are actually allowed.

\TABLE{
\centering
\begin{tabular}{|l||p{9.5cm}|} \hline
& {R/Q-charge Conservation} \\ \hline \hline
Factorizable,  prime & $\sum_\alpha R^i_\alpha = 1 \mod \N^i$, \hspace{0.3cm}$i=1,2,3$   \newline with $R^i_\alpha = q_{sh\,\alpha}^i-\calN_{L\,\alpha}^i +\bar\calN_{L\,\alpha}^i$\\  \hline
Factorizable, non-prime &  $\sum_\alpha R^j_\alpha = 1 \mod \N^j$, \hspace{0.3cm}$j=$ prime planes  \newline with $R^j_\alpha = q_{sh\,\alpha}^j-\calN_{L\,\alpha}^j +\bar\calN_{L\,\alpha}^j$, \newline $\sum_\alpha R_\alpha = (\sum_{i\neq j} \N v^i) \mod \N$ \newline with $R_\alpha = \sum_{i\neq j} \N v^i ( q_{sh\,\alpha}^i-\calN_{L\,\alpha}^i +\bar\calN_{L\,\alpha}^i) + \N\gamma_\alpha$\\ \hline
Non-factorizable,  prime & none  \\  \hline
Non-factorizable,  non-prime & $\sum_\alpha Q_\alpha = 0 \mod \N$ \newline with $Q_\alpha = \sum_{i=1}^3 \N v^i(q_{sh\,\alpha}^i-\calN_{L\,\alpha}^i +\bar\calN_{L\,\alpha}^i) + \N\gamma_\alpha$\,,  \newline   $\sum_\alpha R^i_\alpha = 1 \mod \N^i/n$,  \hspace{0.3cm} for $(\theta_i)^n$ automorphism \newline with $R_\alpha^i = q_{sh\,\alpha}^i-\calN_{L\,\alpha}^i +\bar\calN_{L\,\alpha}^i$ \\  \hline
\end{tabular}

\vspace{0.5cm}

\begin{tabular}{|c||p{9.5cm}|} \hline
& Rule 4 for all twisted sectors at same fixed point\\ \hline \hline
Factorizable, prime & $\calN_L^i -\bar\calN_L^i-\bar\calN_R^i=0 \mod 2\N^i$, \hspace{0.3cm}$i=1,2,3$  \\  \hline
Factorizable, non-prime & $\calN_L^i -\bar\calN_L^i-\bar\calN_R^i=0 \mod \N^i$, \hspace{0.3cm}$i=1,2,3$ \\ \hline
Non-factorizable, prime & $\sum_{i=1}^3(\calN_L^i -\bar\calN_L^i-\bar\calN_R^i)=0 \mod 2$   \\  \hline
Non-factorizable, non-prime & $\sum_{i=1}^3\N v^i(\calN_L^i -\bar\calN_L^i-\bar\calN_R^i)=0 \mod \N$ \newline  $\sum_{i=1}^3(\calN_L^i -\bar\calN_L^i-\bar\calN_R^i)=0 \mod 2$     \\  \hline
\end{tabular}

\vspace{0.5cm}

\begin{tabular}{|c||p{9.5cm}|} \hline
& Rule 6 with $\K=lcm(l_1,\dots,l_\L)$   \\ \hline \hline
Factorizable, prime & none  \\  \hline
Factorizable, non-prime & $\calN_L^i -\bar\calN_L^i-\bar\calN_R^i=0 \mod \N^i/\K$, 
\hspace{0.3cm}for two non-prime $i$ \\ \hline
Non-factorizable, prime & none \\  \hline
Non-factorizable, non-prime & none
\\  \hline
\end{tabular}

\caption{R/Q-charge conservation, Rule 4 and Rule 6 for the various $T^6/\Z_\N$ orbifolds, with $T^6$ factorizable or non-factorizable, and $\N$ prime or non-prime.  We write the order of the orbifold twist in plane $i$ as $\N^i$.  For $\N$ non-prime, $\N^i$ may be prime or non-prime.  We use $\calN_{L\, \alpha}, \bar\calN_{L\, \alpha}$ for the oscillator numbers of each state, and $\calN_{L}, \bar\calN_{L}, \bar\calN_R$ for the total number of oscillators in the correlation function.  The fixed points to which the twisted states are attached satisfy $\theta^{l_\alpha} f_\alpha=f_\alpha + \lambda$, with $l_\alpha$ the smallest such integer, and $lcm$ stands for lowest common multiple.  
Untwisted states may also participate in the couplings.
The rules for the partially factorizable orbifolds are similar, and can be found explicitly in the text.}
\label{T:rules}
}

The question that comes to the mind of all Orbifolders is now: what is the relevance of the string selection rules for phenomenology?  
\TABLE{
\centering
\begin{tabular}{|c||c|c|c|} \hline 
& Used Rules & Correct R-charges & All Rules \\ \hline \hline
Number of couplings &96 & 156 & 132 \\ \hline
\end{tabular}
\caption{Comparison of the number of allowed trilinear couplings between twisted fields in a $T^6/Z_{6-II}$ model, applying the presently used selection rules (gauge invariance, the space group selection rule and R-charge conservation) with old R-charges, the presently used selection rules with correct R-charges, and the
presently used selection rules with correct R-charges plus Rules 4, 5 and 6 (Rule 6 is in fact not independent for $T^6/Z_{6-II}$).  The gauge shift vector defining the model is $V^I=(\frac{1}{3}, -\frac{1}{2}, -\frac{1}{2}, 0, 0, 0, 0, 0, \frac{1}{2}, -\frac{1}{6}, -\frac{1}{2}, -\frac{1}{2}, -\frac{1}{2}, -\frac{1}{2}, -\frac{1}{2}, \frac{1}{2})$, with gauge group $SO(10) \times SU(2) \times SU(2) \times SO(14)\times U(1)^2$.} \label{T:Z6II}
}
We expect the charge conservation laws to be associated with symmetries in the low energy effective field theory.
However, such an interpretation seems more challenging for Rules 4 and 6, as they apply only to some kinds of couplings, depending on the relative properties of the participating twisted strings.  Thus, intriguingly, it would seem that Rules 4 and 6 correspond to ``stringy miracles'' from the point of view of a 4D observer.  The same can be said for Rule 5.

So far, we have only considered models without discrete Wilson lines, whereas all promising models use discrete Wilson lines to reduce the gauge group and the number of generations.  In the presence of discrete Wilson lines, new so-called shift $\gamma$-phases appear \cite{patrick, saulsthesis, jonas}.  It would be important to work out their role in the string couplings. 

Meanwhile, the selection rules we have computed are for couplings in the holomorphic superpotential of the low energy effective field theory.  After moduli stabilization, an effective superpotential emerges, which also has contributions from the K\"ahler potential.  
In \cite{MR, KPRZ}, one can find a special case of an effectively holomorphic matter coupling that is allowed in the effective superpotential, but forbidden by Rules 4 and 5 in the truely holomorphic superpotential.  
Without a better understanding of the K\"ahler potential \cite{DKL}, we cannot write down the allowed couplings in the effective superpotential.

Another interesting issue is to study anomalies for 
the R-symmetries that we have found. 
Such a study on R-symmetry anomalies has been performed for the standard
R-charge conservation law in factorizable orbifolds in
\cite{anomalies}.
It would be important to reconsider such studies 
taking into account our results such as the
inclusion of the $\gamma$ phases.  We would also like to understand further the role of selection rules in the matching between orbifold compactifications and their Calabi-Yau blowups \cite{nana, toffi}.

We close our discussion by writing down an algorithmn to compute the allowed L-point couplings for a $\Z_\N$ orbifold.  We present here the selection rules for the factorizable orbifolds, but the non-factorizable case follows analogously from the main part of the paper and Table \ref{T:rules}.  The rules can be applied as follows:
\begin{enumerate}[1.]

\item {\it Gauge invariance:} $\sum_\alpha p_{sh} = 0$.

\item {\it Space group selection rule:} $\prod_\alpha [(\theta^{k_\alpha},\lambda_\alpha)] = ({\mathbbm 1},0)$.

\item {\it H-momentum conservation:} This does not restrict the allowed couplings, but allows one to determine the right-moving oscillator numbers as $\bar\calN_R^i = \sum_\alpha q_{sh \alpha}^i - 1$.

\item {\it Rule 5} or {\it forbidden instanton selection rule:} Check whether holomorphic and anti-holomorphic instantons in the $i$-th plane are allowed.  Non-trivial holomorphic solutions exist if and only if 
$1 + \sum_\alpha (-1 + \k^i_\alpha) < 0$ (where $0 < \k^i_\alpha \leq 1$).  Non-trivial anti-holomorphic solutions exist if and only if $1 
+
\sum_\alpha (- \k_\alpha^i) <0$ (where $0 \leq \k^i_\alpha < 1$).  If both holomorphic and anti-holomorphic solutions are vanishing, then we require 
${\cal N}_L^i = \bar{\cal N}_L^i + \bar{\cal N}_R^i$.  If  holomorphic instantons are allowed, but anti-holomorphic instantons are vanishing, then 
${\cal N}_L^i \geq \bar{\cal N}_L^i$.  If instead only anti-holomorphic instantons are allowed, then ${\cal N}_L^i \leq \bar{\cal N}_L^i+\bar{\cal N}_R^i$.

\item {\it R-charge conservation:}  Applies when instanton solutions are allowed in the plane $i$.  For planes $j$ with prime ordered twist $\N^j$, we have $\calN_L^j - \bar\calN_L^j - \bar\calN_R^j = 0 \mod \N^j$.  For non-prime planes, we have $\sum_{i\neq j} \N v^i(\calN_L^i - \bar\calN_L^i - \bar\calN_R^i) - \N \sum_\alpha \gamma_\alpha = 0 \mod \N$.

{\item} {\it Rule 6} or {\it coset vector selection rule:}  Applies when two planes $i\neq j$ are non-prime, and when instanton solutions are allowed in the non-prime planes.  For couplings with $\K < \N$, for $\K$ the lowest common multiple of the twisted sectors (or their conjugates), we have $\calN_L^i - \bar\calN_L^i - \bar\calN_R^i = 0 \mod \frac{\N^i}{\K}$. 

{\item} {\it Rule 4} or {\it torus lattice selection rule:}  Applies when instanton solutions are allowed in the planes $i$.  When all twisted sectors are at the same fixed point in plane $i$ for every auxiliary coupling, we have $\calN_L^i - \bar\calN_L^i - \bar\calN_R^i = 0 \mod \M^i$ where $\M^i$ is the order of the torus lattice automorphism group in plane $i$.

\end{enumerate}

\section*{Acknowledgements}  
We would like to thank M.~Blaszczyk, S.~F\"orste, C.~L\"udeling, P.~Oehlmann, N.~Pagani, S.~Ramos-S\'anchez, F.~R\"uhle and P.~Vaudrevange for useful discussions. 
We also thank the organizers of the 4th Bethe Center Workshop and Bethe Forum, Bad Honnef, Germany for kind hospitality while part of this work was done. 
D.~M.~P.~and M.~S.~would also like to thank the Simons Center for Geometry and Physics and the organizers of the Summer School on String Phenomenology in Stony Brook (NY) where part of this work was done.
N.~G.~C.~B.~is partially supported by ``Centro de Aplicaciones Tecnol\'ogicas y Desarrollo Nuclear'' (CEADEN), ``Proyecto Nacional de Ciencias B\'asicas Part\'iculas y Campos'' (CITMA, Cuba) and  the European Commission under the contract PITN-GA-2009-237920 during her stay at CERN. T.~K.~is supported in part by  the Grant-in-Aid for the Global COE  Program ``The Next Generation of Physics, Spun from Universality and Emergence" from the Ministry of Education, Culture, Sports, Science and Technology of Japan. S.~L.~P.~is funded by Deutsche Forschungsgemeinschaft inside the ``Graduiertenkolleg GRK 1463''.
The work of N.~G.~C.~B., D.~M.~P.~and M.~S.~was partially supported by the SFB-Tansregio TR33
``The Dark Universe" (Deutsche Forschungsgemeinschaft) and
the European Union 7th network program ``Unification in the
LHC era" (PITN-GA-2009-237920).

\appendix
\section{A Classification of Orbifold Automorphisms} \label{auto}

In this appendix we present a more general discussion of the orbifold automorphisms.  As we saw in the paper, a crucial point is that the instanton solutions are related to the fixed points by means of the coset vectors. 
We are thus interested in symmetries that respect the fixed points of the orbifold.  
Since the symmetries of the fixed points can be related to the lattice automorphisms, we take this group as the starting point of our exploration. This discrete group is a subgroup of the Lorentz group in the compact six dimensional space. Given that the Lorentz symmetry is the only one which makes a distinction between bosons and fermions, the members of $\text{Aut}(\Lambda)$ which happen to survive the orbifold identifications seem suitable to explain the presence of discrete $R$-symmetries in the low energy effective theory. 

\FIGURE{
\includegraphics[width=11.25cm]{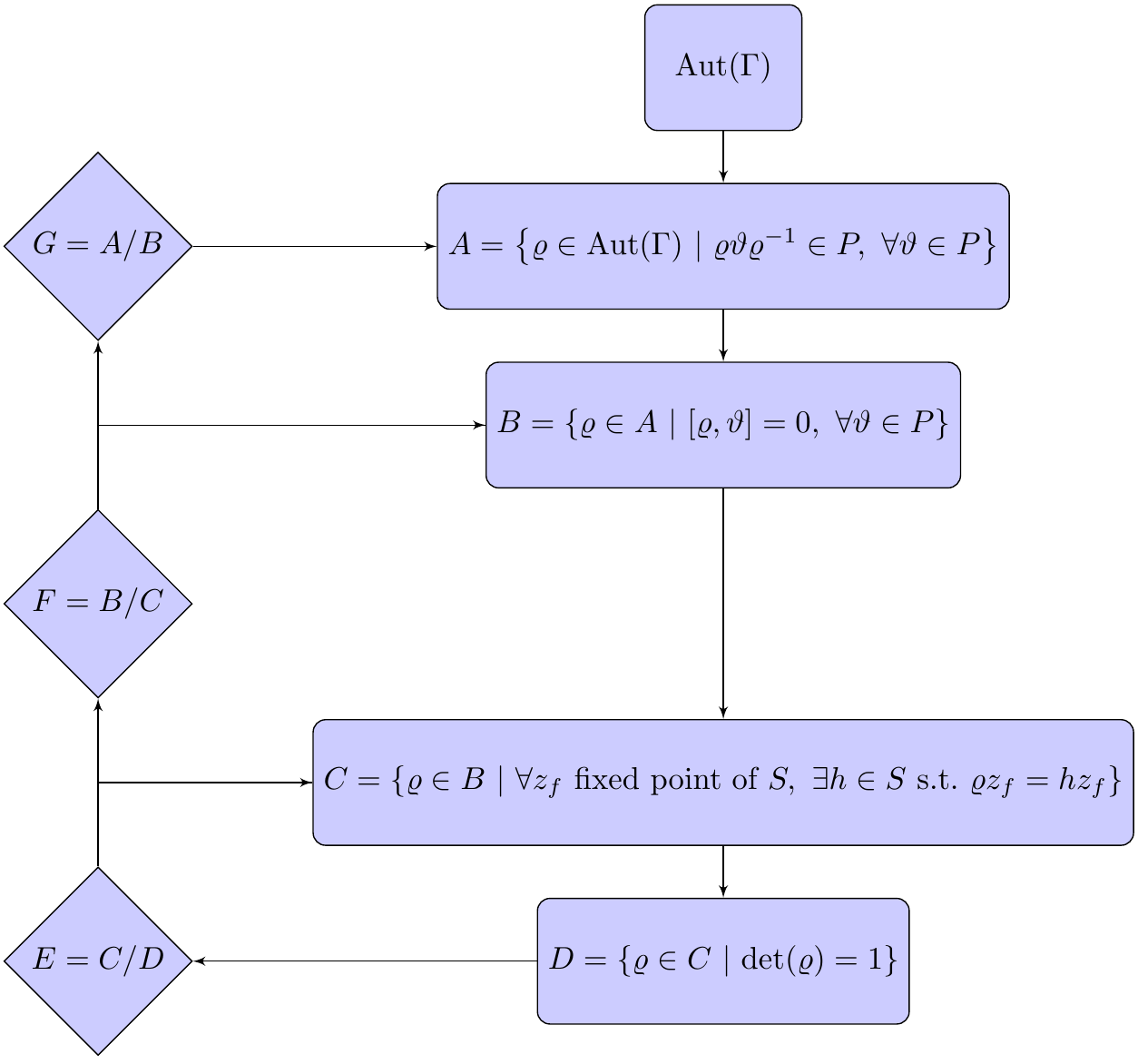}
\caption{The subgroup $A$ of the automorphism group of the six dimensional lattice allows for a decomposition into the subgroups $B$, $C$ and $D$, as defined in eqs.~\eqref{GroupB}, \eqref{GroupC} and \eqref{GroupD}. Provided the normalcy relations between them, one can construct the quotients $E$, $F$ and $G$ which allow for a simpler interpretation given their reduced number of elements.}
\label{taxonomy}}

In order to classify the elements of $\text{Aut}(\Lambda)$, we define certain subgroups, as displayed in Figure \ref{taxonomy}. The biggest subgroup, $A \subset \text{Aut}(\Lambda)$ is the group of automorphisms that respect the point group $P$,
\begin{equation}
A=\left\{\varrho\in\text{Aut}(\Gamma)~|~\varrho\, \theta^k\varrho^{-1}\in P,~\forall \,\,\theta^k\in P\right\}\,.
\end{equation}
Clearly any vector in the compact dimensions will transform under the elements of $\text{Aut}(\Lambda)$. Consequently the action of any $\varrho \in A$ on a space group element $h=(\theta^k,\lambda)\in S$ is given by
\begin{equation}
\varrho(h)=(\varrho\,\theta^k\varrho^{-1},\varrho\lambda)\,.
\end{equation}
Note that $A$ is constructed in such a way that it preserves the structure of the conjugacy classes.  That is, given two space group elements $g_1,g_2$ which belong to the same conjugacy class $[g]$, any transformation $\varrho\in A$ will preserve the network of identifications, i.e. $\varrho(g_1)\sim\varrho(g_2)\in [\varrho(g)]$. 
We define further the subgroups
\begin{align}
B&=\left\{\varrho\in A~|~[\varrho,\theta^k]=0,~\forall \,\theta^k\in P\right\}\,,\label{GroupB}\\
C&=\left\{\varrho\in B~|~\forall \,z_f\text{ fixed point of } S,~\exists \,h \in S~\mathrm{s.t.}~\varrho z_f=hz_f\right\}\,,\label{GroupC}\\
D&=\left\{\varrho\in C~|~\text{det}(\varrho)=1\right\}\,,\label{GroupD}
\end{align}
where $B$ is the subgroup of symmetries which map between conjugacy classes of the same twisted sector. $C$ is defined as the subgroup of automorphisms which preserve all the conjugacy classes of the space group, i.e.~$[\varrho(g)]=[g]$, and $D$ contains all elements in $C$ which belong to $SO(6)$. It is easy to show, that these subgroups fulfill
\begin{equation}
D\lhd C \lhd B \lhd A\,,
\label{normalcychain}
\end{equation}
and hence it makes sense to define the corresponding quotient groups $E=C/D$, $F=B/C$ and $G=A/B$.

The elements in $G=A/B$ are symmetries which exchange space group elements of different twisted sectors, whereas the elements of $F=B/C$ map between inequivalent fixed points within the same twisted sector. The group $E=C/D$ contains all reflections in $O(6)$ which commute with the point group and map all conjugacy classes of the space group to themselves. These quotient groups are very interesting objects to study but the discussion of their implications in the string theory is beyond the scope of this work. In the main text we restricted to elements of group $D$ because they commute with the point group, leave all fixed point conjugacy classes invariant and can be written in terms of the Cartan generators of $SO(6)$.

\end{document}